\documentclass[twocolumn]{aastex63}
\usepackage{hyperref}

\newcommand\mgii{\ion{Mg}{2}}

\newcommand\civ{\ion{C}{4}}
\newcommand\ciii{\ion{C}{3}}
\newcommand\cii{\ion{C}{2}}
\newcommand\ciistar{\ion{C}{2}*}
\newcommand\siii{\ion{Si}{2}}
\newcommand\siiii{\ion{Si}{3}}
\newcommand\siiv{\ion{Si}{4}}
\newcommand\oi{\ion{O}{1}}
\newcommand\ovi{\ion{O}{6}}
\newcommand\sii{\ion{Si}{1}}
\newcommand\ci{\ion{C}{1}}
\newcommand\mgi{\ion{Mg}{1}}
\newcommand\alii{\ion{Al}{2}}
\newcommand\aliii{\ion{Al}{3}}
\newcommand\feii{\ion{Fe}{2}}
\newcommand\hi{\ion{H}{1}}

\newcommand\lya{Ly $\alpha$~}
\newcommand\kms{km s$^{-1}$}
\newcommand\nhi{$N_{\rm HI}$}

\submitjournal{{\it The Astrophysical Journal}}

\shorttitle{Abundances and Ionization of a $z=6.84$ Quasar Absorber}
\shortauthors{Simcoe et al.}

\begin{document}

\title{Interstellar and Circumgalactic Properties of an Unseen $z=6.84$ Galaxy:\\
  Abundances, Ionization, and Heating in the Earliest Known Quasar Absorber}

\correspondingauthor{Robert A. Simcoe}
\email{simcoe@space.mit.edu}

\author{Robert A. Simcoe}
\affiliation{MIT Kavli Institute for Astrophysics and Space Research, 241 Ronald McNair Bldg., Cambridge MA 02139}

\author{Masafusa Onoue}
\affiliation{Max Planck Institute for Astronomy, K\"{o}nigstuhl 17, 69117, Heidelberg, Germany}

\author{Anna-Christina Eilers}\thanks{NASA Hubble Fellow}
\affiliation{MIT Kavli Institute for Astrophysics and Space Research, 241 Ronald McNair Bldg., Cambridge MA 02139}

\author{Eduardo Ba\~nados}
\affiliation{Max Planck Institute for Astronomy, K\"{o}nigstuhl 17, 69117, Heidelberg, Germany}

\author{Thomas J. Cooper}
\affiliation{Carnegie Observatories, 813 Santa Barbara St., Pasadena CA 91101}

\author{G\'{a}bor F\H{u}r\'{e}sz}
\affiliation{MIT Kavli Institute for Astrophysics and Space Research, 241 Ronald McNair Bldg., Cambridge MA 02139}

\author{Joseph F. Hennawi}
\affiliation{Department of Physics, University of California, Santa Barbara, CA, 93106}

\author{Bram Venemans}
\affiliation{Max Planck Institute for Astronomy, K\"{o}nigstuhl 17, 69117, Heidelberg, Germany}

\begin{abstract}

We analyze relative abundances and ionization conditions in a strong
absorption system at $z=6.84$, seen in the spectrum of the $z=7.54$
background quasar ULAS J134208.10+092838.61. Singly ionized C, Si, Fe,
Mg, and Al measurements are consistent with a warm neutral medium that
is metal-poor but not chemically pristine.  Firm non-detections of
\civ ~and \siiv ~imply that any warm ionized phase of the IGM or CGM
has not yet been enriched past the ultra-metal-poor regime ($<0.001Z_\odot$), unlike
lower redshift DLAs where these lines are nearly ubiquitous.  Relative
abundances of the heavy elements 794 Myr after the Big Bang resemble
those of metal-poor damped Lyman Alpha systems at intermediate
redshift and Milky Way halo stars, and show no evidence of enhanced
[$\alpha$/Fe], [C/Fe] or other signatures of yields dominated by
massive stars.  A detection of the \ciistar ~fine structure line
reveals local sources of excitation from heating, beyond the level of
photo-excitation supplied by the CMB.  We estimate the total and
[\cii] cooling rates, balancing against ISM heating sources to develop
an heuristic two-phase model of the neutral medium.  The implied
heating requires a surface density of star formation slightly
exceeding that of the Milky Way but not at the level of a strong
starburst.  For a typical (assumed) \nhi=$10^{20.6}$, an abundance of
[Fe/H]$=-2.2$ matches the columns of species in the neutral phase.  To
remain undetected in \civ, a warm ionized phase would either need much
lower [C/H]$<-4.2$ over an absorption path of 1 kpc, or else a very
small absorption path (a few pc).  While still speculative, these
results suggest a significant reduction in heavy element enrichment outside
of neutral star forming regions of the ISM, as would be expected in
early stages of galactic chemical evolution.

\end{abstract}


\section{Introduction} \label{sec:intro}

As increasing numbers of quasars are discovered at $z>6.5$
\citep{mortlock,yang2020,pisco,wang2018,reed2019,venemans2015,mazzucchelli2017,wang2019},
their spectra provide pathlength for detecting randomly intervening
absorption systems via heavy-element lines.  Although the incidence
rate of highly-ionized \civ ~systems declines toward higher redshift
\citep{2013MNRAS.435.1198D,2011ApJ...743...21S}, the frequency of
$W_r<1$\AA ~\mgii ~absorbers and other singly-ionized lines remains nearly
constant to the highest redshifts yet observed
\citep[though stronger systems do evolve;][]{2011ApJ...735...93B,chen_fire_mgii,2012ApJ...761..112M,2020arXiv201011432Z}.  Few
$z>6$ heavy element absorbers have been observed at finer than
echellette spectral resolution ($\sim 50$ \kms), but their high \mgii,
\cii, and \siii~ column densities and non-evolving number counts
suggest an association with early circumgalactic or even interstellar
gas rather than the widespread intergalactic medium.

From $0<z<5.5$, the most prominent signpost of circumgalactic gas is
the \civ ~doublet, which arises in matter that has density $n\sim
10^{-4}-10^{-3}$ cm$^{-3}$ and is at or near equilibrium with the
ambient UV radiation background.  New evidence is accumulating that
before $z\sim 6$, observational signatures of this highly ionized
phase of the CGM are diminishing \citep{cooper2019,2011ApJ...735...93B}.

The most distant absorption system in the literature is seen towards
the $z=7.54$ quasar ULAS J134208.10+092838.61 \citep[hereafter ULAS
  J1342,][]{pisco}. Besides being the redshift record-holder, this
absorber also presents strong characteristics of low-ionization
systems that are rare at $z<6$ but appear to dominate the population
at high redshift.  It is one of several such examples with $N_{\rm
  CII}\gg N_{\rm CIV}$ identified by \citet{cooper2019}.  They argued
that a softening of the UV background is responsible for the
disappearance of the highly ionized phase within this population as a
whole, based on artificially softened toy models of the background
spectrum, modified from \citet{HM2012}.

\begin{figure*}
  \epsscale{1.0}
  \includegraphics[scale=0.7]{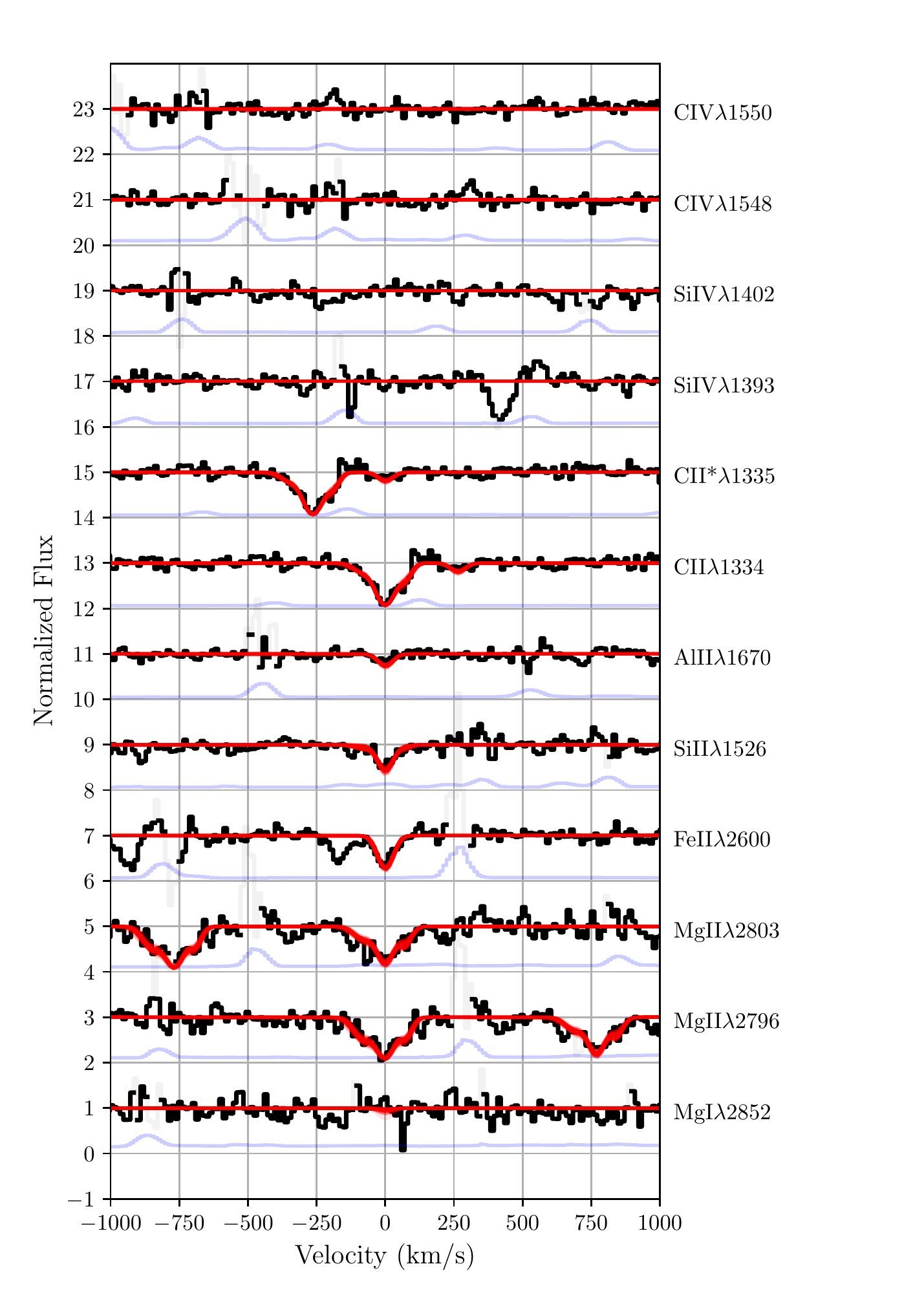}
  \hspace{-1.2cm}
  \includegraphics[scale=0.7]{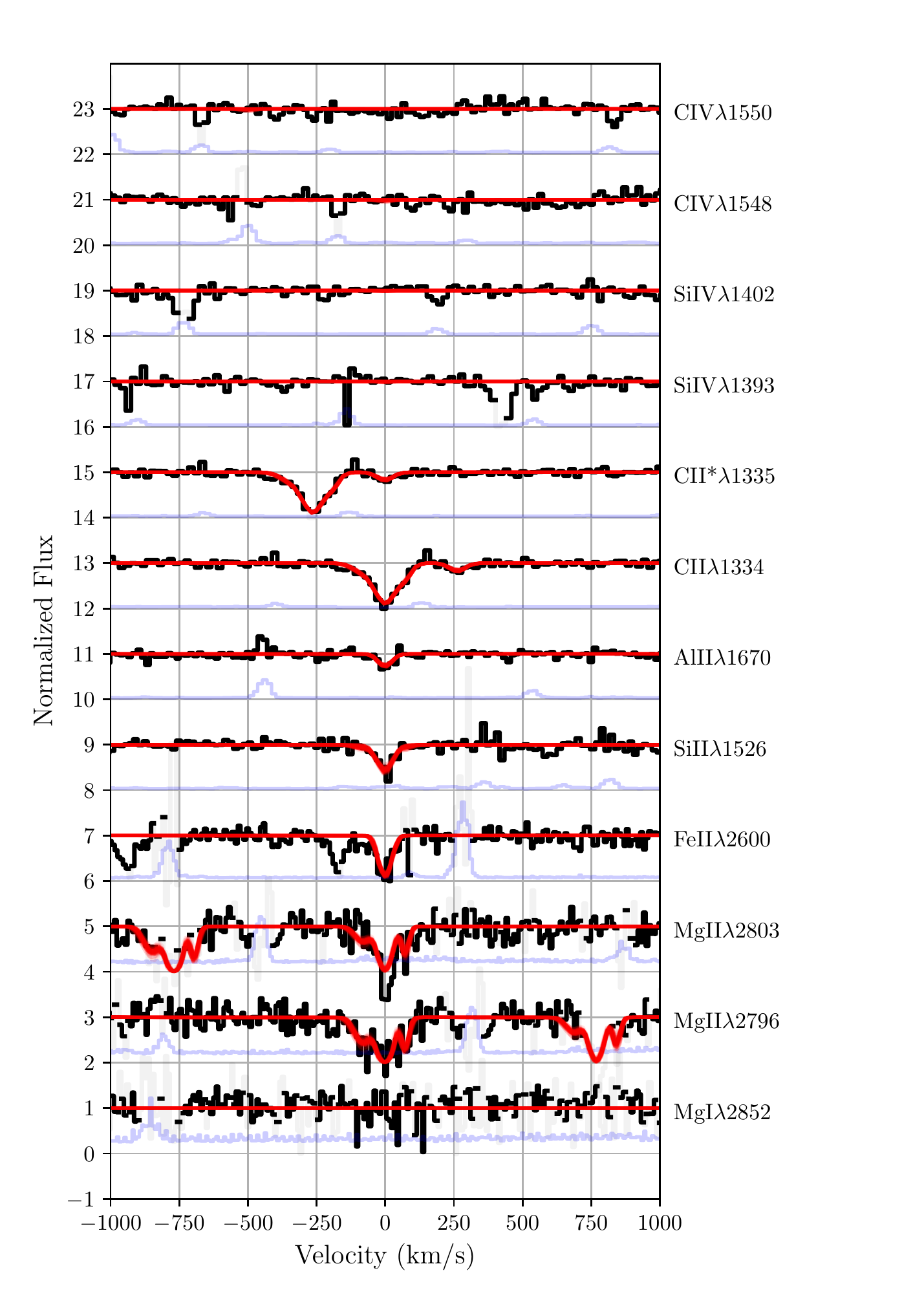}
  \caption{Velocity stack of key ions centered at $z=6.84314$, as
    observed in a 10-hour FIRE spectrum (left panel) and a 23-hour
    XShooter spectrum (right panel).  Data are shown in black bins,
    with 75 absorption profiles drawn randomly from the posterior
    model distribution (red).  Light gray curves show $1\sigma$ error
    vector for spectral data. Regions of high Poisson noise near sky
    lines are shown with reduced line transparency (appearing grayed)
    in the spectrum for clarity.}
  \label{fig:velplot}
\end{figure*}

Here we present a detailed analysis of this exemplar absorber,
including more sophisticated fits to observed column densities (or
their upper limits) using a new Markov Chain Monte Carlo
metal-absorption fitter, and evaluation of the derived relative
abundances of C, Fe, Si, Al, and Mg.  These measurements include
detection of the \ciistar ~excited fine structure line from ions in
the $^2P_{3/2}$ state, which we use to develop an interstellar
heating/cooling model of a two-phase ISM, inspired by earlier work of
\cite{wolfe2003}.  We then examine the ratios of \civ/\cii ~and
\siiv/\siii ~compared to lower redshift examples, and discuss
implications for the ISM's ionized phase, as well as the role of the
UV background radiation field at $z=6.84$.  Throughout analysis we
adopt cosmological parameters $H_0=69.6$ km s$^{-1}$ Mpc$^{-1}$,
$\Omega_{\rm M}=0.286$ and $\Omega_\Lambda=0.714$.

\section{Observations} \label{sec:observations}

Our analysis uses FIRE \citep{2013PASP..125..270S} and XShooter
\citep{xshooter} spectra of ULAS J1342 obtained over multiple
different observing runs. The first FIRE run from January 2017 was for
the discovery spectrum reported in \citet{pisco}.  Additional data
were obtained in April 2018 and added to the original to increase
signal-to-noise ratio.  The total effective exposure time for the
combined FIRE data is 10 hours. All data were taken in FIRE's echelle mode
with a $0.^{\prime\prime}6$ slit, yielding R=6000 from
$0.8<\lambda<2.5\mu$m.  Reductions were made with the {\tt firehose}
package and corrected for telluric absorption using A0V standard
stars.

To improve sensitivity we include an independently-obtained XShooter
spectrum of J1342, combined from two programs (PID: P0100.A-0898(A)
and P098.B-0537(A)) with integration time 22.67 hours, initially
presented in \citet{2020arXiv201006902S}.  These data were reduced
using the {\tt PypeIt} reduction package \citep{pypeit}, and use
XShooter's $0.^{\prime\prime}6$ slit, yielding $R=8100$ spectra over
the same wavelength range.  The XShooter spectra were corrected for
telluric absorption using a joint fit of a telluric model and quasar
continuum PCA, in contrast to the FIRE data which employed A0V stars.

Rather than co-adding spectra from different instruments and different
resolution, we performed the analysis below using a single forward
physical model to fit pixel data from both spectra simultaneously,
with appropriate instrumental convolution kernels.  This preserves
independence of the data sets but requires a match between model and
data in both cases.  Though both spectra were used in every portion of
the analysis, the deep XShooter spectrum had higher SNR at $\lambda <
1.2 \mu$m, while the FIRE data had higher SNR at $\lambda > 1.8\mu$m
(they are comparable in the $H$ band). XShooter's contribution was
therefore critical for detection of the \ciistar ~line, while FIRE was
needed for proper characterization of \mgii ~and set important limits
on non-detection of \mgi.

Prior to calculating absorption properties, we fit a continuum spline
to normalize the data, with knots initially spaced in 1500 km/s
increments and set to median flux levels with rejection of outlying
pixels.  These were adjusted slightly by hand to remove knots that
were contaminated by strong absorption systems, or add structure
around emission line peaks and other regions where one finds strong
variation on small spatial scales. As with other analyses of narrow
absorption lines, the continuum fit is very smooth across the $\sim
150$ km s$^{-1}$ velocity scale characteristic of the foreground
system.

\section{Absorption Line Fits}

\begin{figure}
  \epsscale{1.45}
  \hspace*{-1cm}\plotone{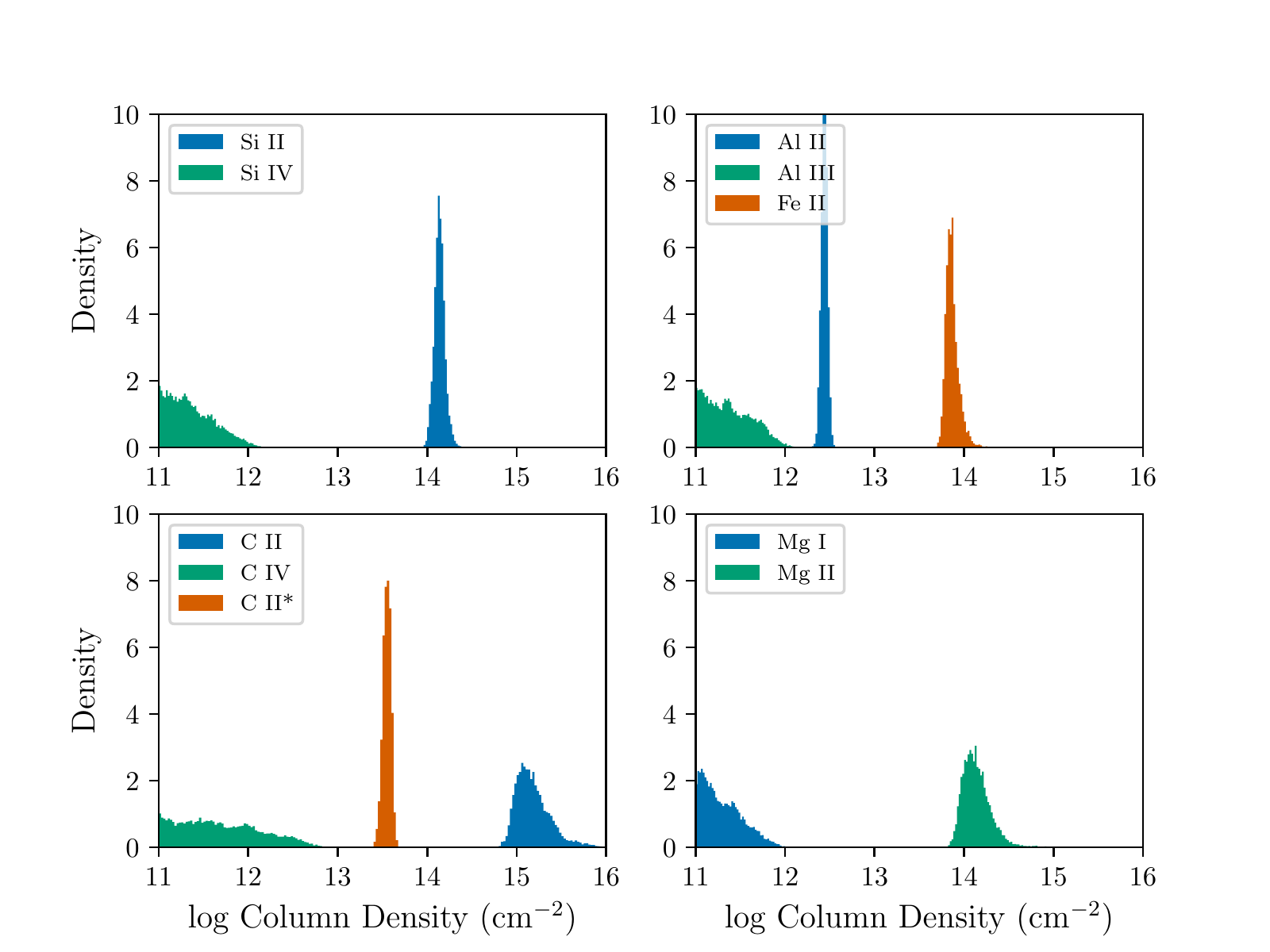}
  \caption{Marginalized posterior distributions of ion column density
    for transitions as listed in Table 1. Non-detections are evident
    for \civ, \siiv, \mgi ~and \aliii, whose posterior distributions
    extend to the minimum value of their (uniform) input priors}
  \label{fig:column_posteriors}
\end{figure}

We developed a new python-based code to fit column densities directly
to the spectral data using Markov Chain Monte Carlo (MCMC) samplers, which
capture degeneracies in the fitting parameters.  This approach also
explores the effect of saturation on the posterior probability
distributions of column density more comprehensively than
curve-of-growth or apparent optical depth \citep{1991ApJ...379..245S}
methods.

The software creates an absorption {\tt Model} class hierarchy
consisting of one or more fitting {\tt Components}, which are
specified by a redshift, a thermal $b$ parameter (which maps to
temperature), and a turbulent $b_{turb}$ parameter.  Each {\tt Component} has
multiple {\tt Ion} children, which are specified by their name
(e.g. \cii) and column density, and a dictionary of {\tt Transitions}.
The {\tt Transitions} correspond to a single line, and are
characterized by rest wavelength, and fundamental atomic data \citep{2003ApJS..149..205M}.

The user builds up a {\tt Model} by entering {\tt Components, Ions},
and {\tt Transitions}, along with their associated priors on $z, N, b$
and (optionally) $b_{turb}$, and appropriate fitting regions, which can be drawn from
multiple spectra using different instruments and resolutions.  The
{\tt Model} class contains methods to construct Voigt profiles
smoothed by the instrumental line spread function, given a vector of
fit parameters. By arranging the model hierarchy in this fashion, one
naturally fits absorption components with multiple ions and column
densities, but a single value for redshift or Doppler parameter, that
is perturbed for all lines together during the fit, or a single $N$
for each {\tt Transition} of an {\tt Ion}.  A model for a single
component with column densities measured for 11 distinct species
contains 14 fit parameters.  As a default we fit regions within
$\Delta v = \pm 125$ ~\kms ~of the transition center, with some
adjustments for individual noisy regions.

The input model is passed to the {\tt emcee} sampler package \citep{emcee}
to run MCMC chains and evaluate the posterior distributions of {\tt
  Model} parameters.  By default the model is set up with flat priors
on all redshifts, turbulent $b$ parameters, temperatures and column
densities with $11<\log (N)<17$, $3< \log T < 5$, and $2<b<100$ \kms.

\begin{deluxetable*}{lllll}
  \centerwidetable
  \tablewidth{0pt}
\tablecaption{MCMC Voigt Profile Fitting Results}
\label{tab:vpfits}
\tablecolumns{5}
\tablehead{\colhead{$z$} & \colhead{$b_{turb}$ (km s$^{-1}$)} & \colhead{$T$(K)} & \colhead{Ion} & {$N$(cm$^{-2}$)}}
\startdata
6.84336 [6.84332,6.84341] & 21.8 [17.4,26.2] & 10176 [1213,80700] & \cii & 15.09 [14.86,15.54]\\
                          &                  &                    & \siii & 13.55 [13.47,13.62] \\
                          &                  &                    & \siii & 14.09 [14.00,14.20] \\
                          &                  &                    & \mgii & 13.90 [13.73,14.20] \\
6.84128 [6.84106,6.84148] & 35.6 [28.8,46.2]  & 11267 [1300,77841]  & \cii & 13.67 [13.56,13.77]\\
                          &                  &                    & \siii & 12.96 [11.56,13.23] \\
                          &                  &                    & \mgii & 13.10 [13.00,13.21] \\
6.84529 [6.84519,6.84541] & 7.1 [3.5,12.6]  & 6514 [1169,58562]  & \cii & 13.97 [13.76,14.44]\\
                          &                  &                    & \siii & 12.62 [11.42,13.08] \\
                          &                  &                    & \mgii & 13.42 [13.04,14.19] \\
\hline
Totals:                   &                  &                    & \cii & 15.15 [14.94,15.56] \\
                          &                  &                    & \cii$^{*}$ & 13.55 [13.47,13.62] \\
                          &                  &                    & \civ & $<$12.50  \\
                          &                  &                    & \ci & $<$12.85 \\
                          &                  &                    & \siii & 14.14 [14.04,14.24] \\
                          &                  &                    & \siiv & $<$11.85 \\
                          &                  &                    & \sii & $<$12.63 \\
                          &                  &                    & \mgii & 14.12 [13.93,14.43] \\
                          &                  &                    & \mgi & $<$11.72  \\
                          &                  &                    & \feii & 13.86 [13.77,14.01] \\
                          &                  &                    & \alii & 12.44 [12.38,12.50] \\
                          &                  &                    & \aliii & $<$11.82 
\enddata
\tablecomments{Bracketed values encompass 95\% of MCMC posterior distributions. For non-detections, the upper limits represent exclusions at the 95\% confidence level.}
\end{deluxetable*}

The code yields upper limits on $N$ for non-detected ions,
because the posterior extends from the lower limit of the prior
interval to the maximum $N$ consistent with the spectral data.  The
model parameters for $z$ and $b$ in undetected ions are constrained by
other ions in the same component that have significant detections.
The reported upper limits for non-detections represent the value below
which 95\% of the posterior distribution is found.  As such they are
weakly dependent on the lower bound of the chosen prior interval
(since the posterior extends to that bound for a non-detection), but
our overall scientific conclusions are not sensitive to the
particulars of this selection.

\begin{figure}
  \epsscale{1.45}
  \hspace*{-1cm}\plotone{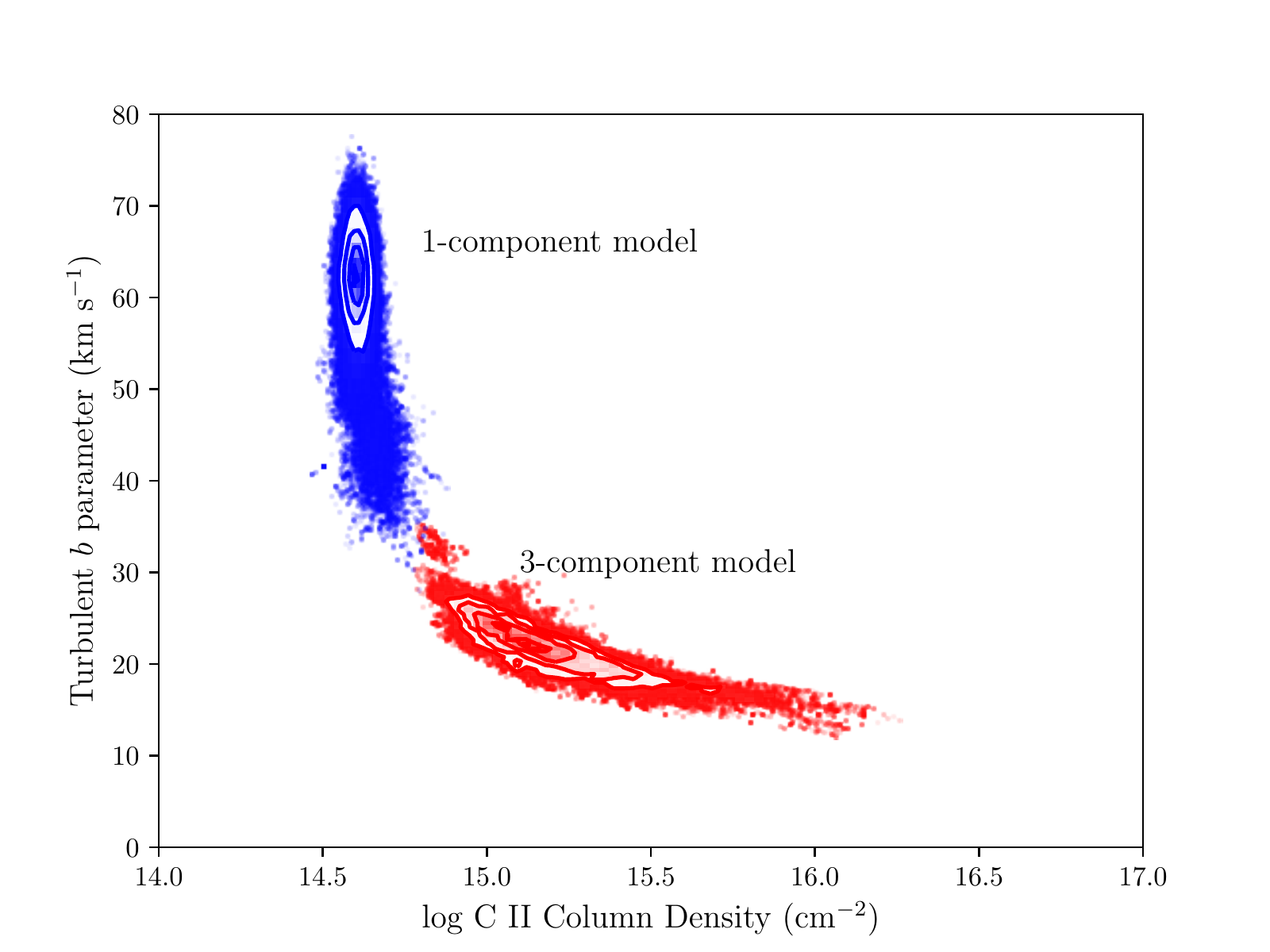}
  \caption{2D posterior showing correlation of \cii ~column density and
    turbulent $b_{turb}$.  The red contours show allowed space for the
    favored 3-component model described in the text, and blue contours
    show allowed space for the alternative 1-component model.
    Although the blue model achieves a comparabe goodness of fit, we
    disfavor it because it requires $b$ much larger than is typically
    observed in lower $z$ heavy-element absorbers observed at
    high-resolution. }
  \label{fig:CII_b_corner}
\end{figure}

When testing the MCMC code, we explored the sensitivity of our derived
column densities to the possible presence of multiple narrow
components, unresolved in our echellette-resolution (50 \kms)
spectra. \citet{cooper2019} noted in HIRES spectra of \oi ~for some
$z>6$ absorbers that there is a very simple velocity structure, with
intrinsic $b<7$ \kms ~(i.e. much smaller than FIRE's resolution
element), and little evidence for complex kinematics---similar to what
was reported by \citet{2011ApJ...735...93B} at $z>6$. Traditional lore
holds that in blended systems, the {\em summed} column density of all
components may be well-constrained, even though the $N$ of the {\em
  individual} components is highly uncertain. Put another way, the
components may trade column density between each other leading to
large individual errors, but their cumulative total is well-specified.
Our tests verified that this was indeed the case for the MCMC fitter.

\section{Results}

\subsection{Column Densities}

Figure \ref{fig:velplot} displays the spectral data from FIRE (left)
and XShooter (Right) with the fitted model profiles overlaid in
red. The model is the same for both panels; the only difference is in
the convolution profile of the line spread function applied to data
from each instrument. A randomly chosen sample of 75 models from the
posterior distribution is plotted, making the width of the red curve a
projection of model uncertainty onto the basis of the data.
Associated fit parameters are listed in Table \ref{tab:vpfits},
reported as median values, with associated 5\% and 95\% confidence
intervals of the posterior distributions.  For undetected ions we
report a 95\% upper bound.

Our preferred model incorporates three absorption components: one
strong line at $z=6.84329$, and weaker flanking components at $\Delta
v = -73$ \kms ~and $+76$ \kms.  All three have best-fit $T\sim
7000-10000$ K and turbulent $b_{turb}\sim 25$ \kms.  These three
components spanning 150 \kms ~are needed to fit broadened \cii ~and
\mgii ~profiles that are well-resolved (spaning three resolution
elements of the spectrograph). The three-component model fits 26
independent parameters, achieving a reduced $\chi^2_\nu=0.94$ per
degree of freedom.  The full posterior parameter distributions are
shown in Figure \ref{fig:column_posteriors}. For \cii, \mgii, and
\siii ~in this figure we show the summed total $N$ of the three
absorption components.

It is possible to achieve a similar quality of fit ($\chi_\nu=1.06$
per degree of freedom) using a one-component model with just 14 fit
parameters, which would seem attractive because of simplicity.
However this model requires a large turbulent broadening ($b_t=62$
\kms) driven by the need to span the resolved width of the \cii ~and \mgii
~profiles.  Figure \ref{fig:CII_b_corner} illustrates this effect by
plotting contours of the 2D $N_{\rm CII}-b$ space explored by the MCMC
walker, with blue contours tracing the one-component fit.

Although the one-component fit yields a narrower marginal distribution
in \cii ~column density and might therefore seem a ``better'' fit,
this is only possible if $55<b<70$ \kms.  Such extremely broad systems
have not been obseved in metal absorbers at lower redshift, and even
at $z>6$ the majority of \oi ~and \cii ~lines observed with HIRES
exhibit low temperatures and quiescent kinematics
\citep{2011ApJ...735...93B,cooper2019}.  We therefore disfavor the
one-component model on physical grounds and concentrate on the
three-component model for remaining analysis.

\begin{figure}
\epsscale{1.3}
  \includegraphics[scale=0.5]{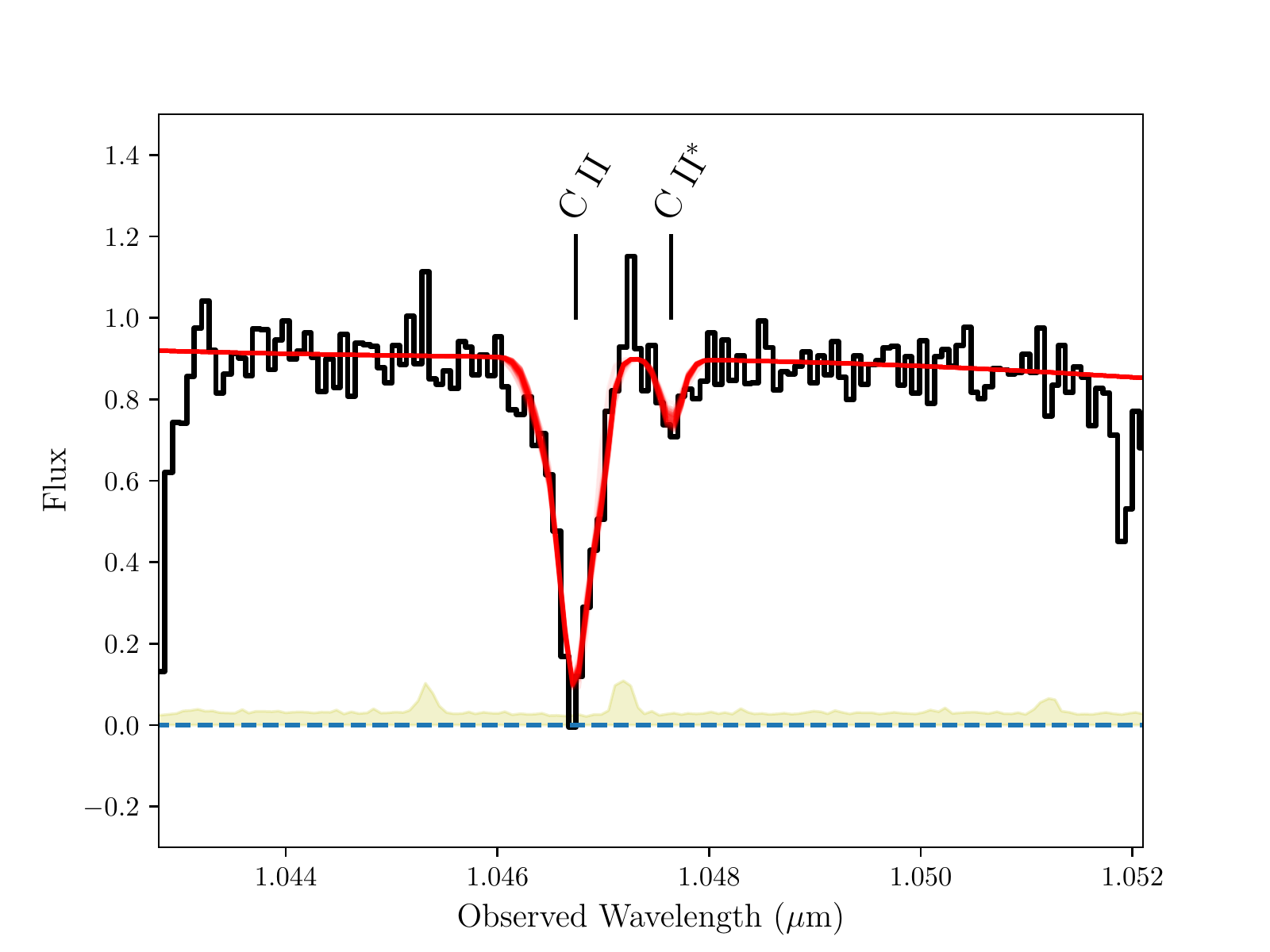}
  \caption{Detailed profile of the \cii$\lambda$1334.5\AA ~and
    \ciistar$\lambda$1335.7\AA ~lines, without normalization by the
    continuum fit.  Red curves indicate the continuum+absorption model
    for 100 random draws of the MCMC posterior.}
  \label{fig:ciistar}
\end{figure}

\begin{figure*}
  \epsscale{1.0}
  \hspace*{-0.5cm}\plotone{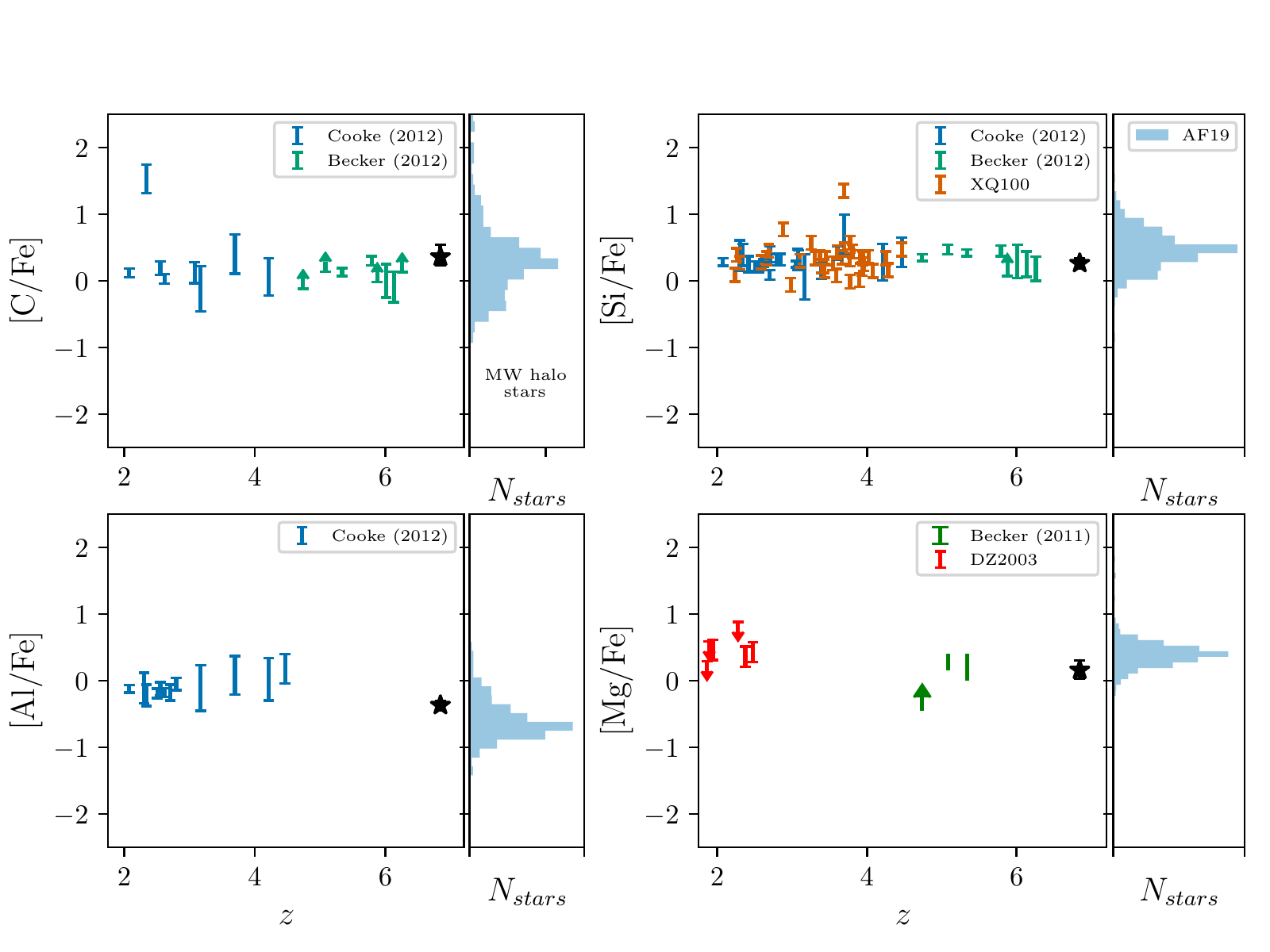}
  \caption{Relative abundance of C, Si, Al and Mg relative to Fe,
    derived from singly ionized species of each element with no
    ionization correction applied.  The new $z=6.84$ measurement is
    illustrated with a black star; literature measurements are
    indicated from lower-redshift quasar absorption line systems
    \citep{2012ApJ...744...91B,2011MNRAS.417.1534C,2003MNRAS.345..447D}.
    The Solar scale is taken from \citet{2009ARA&A..47..481A}.  The
    offset histograms to the right of each panel show the abundance
    distribution of metal-poor stars in the Milky Way's halo and in
    local dwarf galaxies \citep{JINAbase}.  We do not see evidence at
    $z=6.84$ for modified yields as might be expected for Population
    III stars or an evolving IMF. There is also no evidence for a
    change in the [$\alpha$/Fe] ratio as measured with C, Si and Mg,
    even though the Hubble time is just 794 Myr.}
  \label{fig:relative_abundances}
\end{figure*}

In the favored model, \cii ~and \mgii ~arise in mildly saturated
components that are individually unresolved by FIRE, but the overall
complex contains discrete substructures whose kinematic motions {\em
  are} resolved.  For \cii, this leads to a high-end tail in the
column density distribution that can reach as high as $10^{16}$
cm$^{-2}$ (Figures \ref{fig:column_posteriors} and
\ref{fig:CII_b_corner}), but is concentrated between
$10^{14.75}-10^{15.50}$ for $10<b<25$ \kms, which is more typical of
metal absorbers at lower redshift.

We obtain a highly significant detection and measure precise column
density for the \ciistar ~excited state fine structure line, having
verified that the absorption (which is visualized more clearly in
Figure \ref{fig:ciistar}, along with the local continuum fit) is not
from interloping systems at other redshift.  This line is detected in
many DLAs at lower redshift; it is sensitive to local heating in the
neutral phase of interstellar gas, and will be discussed in depth
below.

\begin{figure*}
  \epsscale{0.9}
  \hspace*{-0.5cm}\plotone{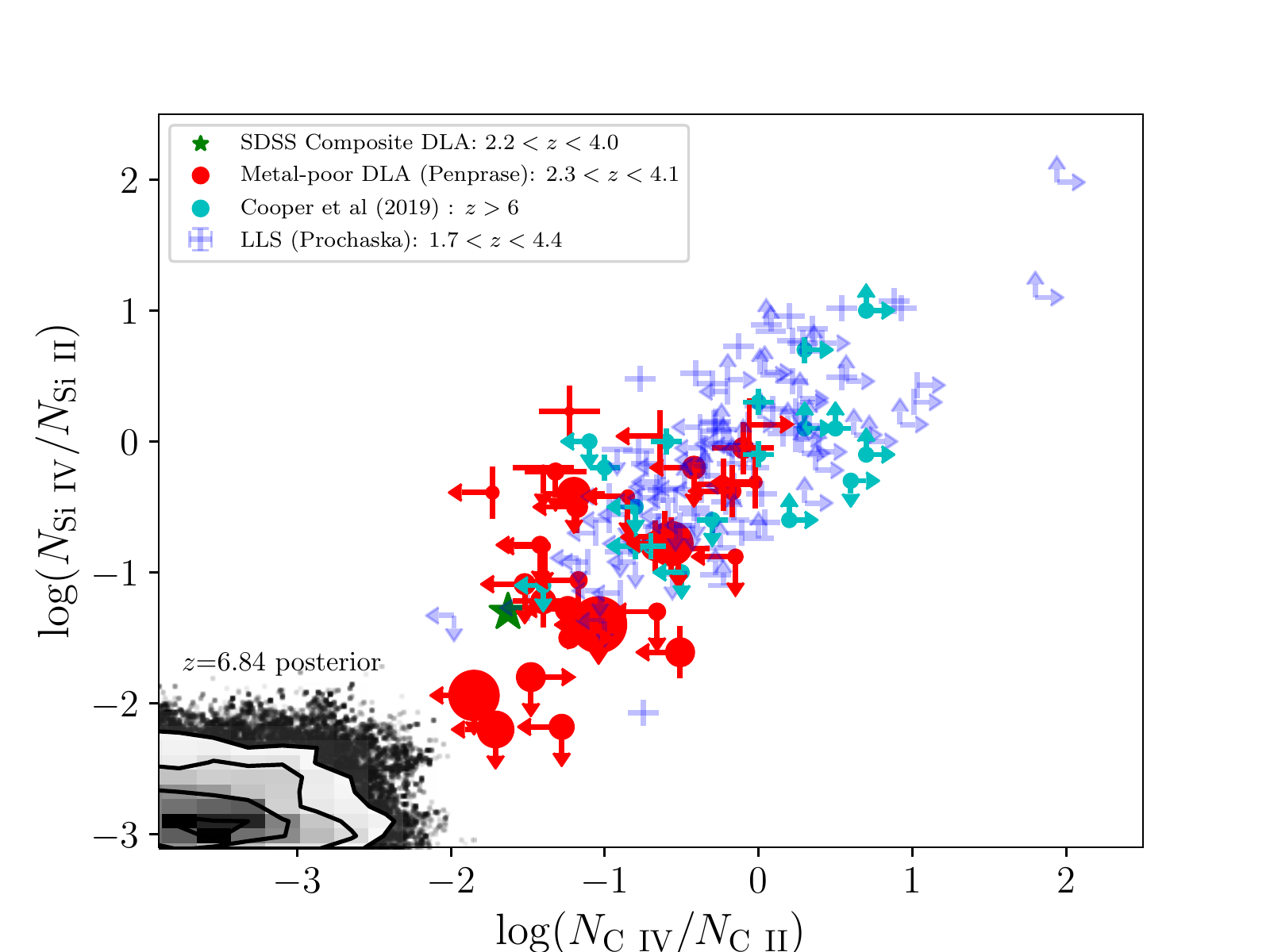}
  \caption{Ratio of triply-ionized to singly-ionized column density
    for C and Si, with increasing degree of ionization toward the
    upper right.  The posterior distribution for the $z=6.84$ absorber
    in ULAS J1342 is shown in the contour/scatter distribution at
    lower left.  Selected samples from the literature are shown with
    colored points as labeled in the legend and referenced in the
    text.  This system has the strongest ionization limits yet
    observed, because of strong detections for \cii ~and \siii~ but
    sensitive upper limits on \civ ~ and \siiv.  Some high-$N_{\rm
      HI}$ but metal-poor DLAs at lower redshift may overlap if more
    sensitive data are obtained.}
  \label{fig:ion_ratios}
\end{figure*}

Remarkably, we do not detect absorption from triply ionized species of
any element --- the upper limits on \civ ~are nearly 2 dex below the
measured \cii ~column.  Unfortunately we cannot observe \ciii
~($\lambda977$\AA) or \siiii ~($\lambda1206$\AA) because of
Gunn-Petersen blanketing in the \lya forest. The presence of
singly-ionized species (and non-detection of \mgi) suggests that this
gas is exposed to radiation in the $E=0.5-1.0$ Ryd band, but that the
radiation field at harder energies (dominated by AGN) is weak compared
to what is seen at lower redshifts.

\subsection{Relative Abundances}

Measurements of absolute heavy element abundances on a hydrogen scale
are impossible at $z>6$ because \hi~ is blended in the forest.
However our detection of \feii ~enables evaluation of {\em relative}
heavy element abundances compared to the Solar pattern, which we
normalize to the photospheric scale of \citet{2009ARA&A..47..481A}.
This constitutes the earliest direct measurement of any abundance,
well within the first Gyr of cosmic history.

Motivated by our non-detection of doubly- or triply-ionized species,
we follow the approach used for abundance analysis of Damped Lyman
Alpha (DLA) absorbers at low redshift, assuming all gas is in the
neutral phase, and no ionization correction is required
\citep{2005ARA&A..43..861W}. This assumption is self-consistent if the
absolute abundance is below roughly [Fe/H]$\le -2.0$; at these levels
the \hi ~column density implied by our \cii ~and \siii ~measurements
would exceed $N_{\rm HI}=10^{20.3}$ cm$^{-2}$ (the canonical DLA
\nhi), using the Solar abundance scale of
\citet{2009ARA&A..47..481A}. We further verified with {\tt cloudy}
calculations that for $N_{\rm HI}=10^{21}$ cm$^{-2}$, the ionization
fractions of our observed transitions were all $\gtrsim 95$\%, and
abundance corrections were at the 0.01-0.02 dex level, far below
random observational errors.  If the absolute abundance is higher,
then the \hi ~column would be lower and an ionization correction may
be required. We revisit this assumption in Section
\ref{sec:ion_ratios}.

\begin{deluxetable}{cr}[b]
  \centerwidetable
  \tablewidth{0pt}
  \tablecaption{Relative Abundance Measurements}
  \tablecolumns{2}
  \tablehead{\colhead{Ratio} & \colhead{\hspace{2.5cm}Median Value\hspace{0.75cm}}}
  \label{tab:relabund}
  \startdata
      {\rm [C/Fe]} & $+0.36 ~[+0.24,+0.54]$ \\
      {\rm [Si/Fe]} & $+0.26 ~[+0.19,+0.33]$ \\
      {\rm [Al/Fe]} & $-0.37 ~[-0.45,-0.31]$ \\
      {\rm [Mg/Fe]} & $+0.16 ~[+0.03,+0.30]$
      \enddata
      \tablecomments{Bracketed values encompass 16/84\% of MCMC posterior distributions.}
\end{deluxetable}

Figure \ref{fig:relative_abundances} shows the abundances of C, Si, Al
and Mg relative to Fe.  In each panel, we show abundances of the same
element measured in other quasar absorption systems at lower redshifts
\citep{2012ApJ...744...91B,2011MNRAS.417.1534C,2003MNRAS.345..447D},
and also distributions for metal-poor stars in the Milky Way and
nearby dwarf galaxies taken from the JINAbase literature compilation
\citep{JINAbase}.  The $z=6.84$ absorption system is shown as a black
star at right.

We do not see any clear evidence at $z=6.84$ of anomalous relative
abundances that would suggest an evolution in heavy element yield
patterns.  In particular we do not observe evidence of a large C
enhancement, as has been seen in some metal-poor halo stars and
suggested as a signature of early metal enrichment \citep{cemp}.

There is tentative evidence of a sub-solar Al abundance, which is
intriguing because metal-poor DLAs at $2<z<4$ appear to follow the
Solar pattern, but metal-poor local stars are also centered around
[Al/Fe]$\sim -0.5$.  Some metal-pool LLS at $2<z<3.5$ also appear
to be deficient in Al \citep{2016ApJ...833..270G}.

Perhaps the most interesting implication is that the [$\alpha$/Fe]
ratio as probed by C, Si, and Mg does not show any marked enhancement
in a cross-section selected gas cloud just $\sim 800$ Myr after the
Big Bang.  This means that Fe production must already be robust even
though enrichment from Type Ia supernovae is unlikely to be fully
underway \citep{2001ApJ...558..351M}. Similar findings have been
extensively documented in broad line regions of $z>6.5$ quasars
\citep{2014ApJ...790..145D,2020arXiv200616268O,mazzucchelli2017,2003ApJ...594L..95B,2003ApJ...596..817D}.
However absorber-derived abundances should be more representative of
the general IGM/CGM than an AGN broad-line region, and they should
also be less susceptible to systematic offsets caused by local
ionizing sources. This extends the findings of
\citet{2012ApJ...744...91B} at $z=6.1$ (922 Myr after the Big Bang)
now to $z=6.84$ (794 Myr).  The absence of strong evidence for unusual
yield patterns suggests that early generations of Population II-like
stars were forming in this neighborhood by $z\gtrsim 10$.

\subsection{Ion Ratios}
\label{sec:ion_ratios}

Ratios of singly- to triply-ionized species of the same element encode
information about ionization, independent of uncertainty of
nucleosynthetic yields or relative abundances.  A striking aspect of
the $z=6.84$ absorber is its very strong \cii ~and \siii ~with
sensitive non-detections of \civ ~and \siiv.  Undetected
high-ionization-potential lines yield only a one-sided constraint on
ionization.  However unlike systems where both \cii ~and \civ ~are
detected, these systems have no ambiguity about whether gas giving
rise to absorption from different species occupies a different phase
region in the temperature-density plane.

In Figure \ref{fig:ion_ratios}, we show with gray 2D contours the
allowed posterior region in the $N_{\rm CIV}/N_{\rm CII}$ vs. $N_{\rm
  SiIV}/N_{\rm SiII}$ plane for the $z=6.84$ absorber.  The ratios are
configured such that more highly ionized systems appear at the upper
right of the plot.  The allowed region for the $z=6.84$ absorber is at
the extreme bottom left---in fact it extends off the corner of the
plot, because of non-detections in \civ ~and \siiv.

We estimate with 95\% confidence that $N_{\rm SiII}:N_{\rm SiIV} >
28:1$ and $N_{\rm CII}:N_{\rm CIV} > 130:1$; the posterior is centered
around a region of much lower ionization, near $N_{\rm CII}:N_{\rm
  CIV}=1200:1$.

To contextualize this result, we show the ratio derived from other
selected samples in the literature that list $N$ for all four ions, at
both low and high redshift.  Light blue points indicate the locus of
Lyman Limit Systems at $z\sim 3$ \citep{2015ApJS..221....2P}; red
points correspond to a \citet{penprase} sample of metal-poor DLAs at
similar redshift.  For the DLA sample, the point size scales with \hi
~column density.  A green star shows the value for a composite spectrum
of DLAs stacked from the SDSS \citep{khare_sdss_dla_composite}.

The absorber studied here has the highest redshift, and also the
lowest ionization ratios yet observed for carbon and silicon.  This is
especially true for the \civ/\cii ~ratio, which has limits 1-2 dex
below the values observed for many LLS and DLAs.

Several points are shown in cyan from \citet{cooper2019}, who measured
a larger sample of low-ionization absorbers at high redshift. Many of
these may also overlap with the system studied here, but they
generally have two-sided upper limits on these ratios that are less
sensitive. Better spectra could place them in the region measured
for our absorber.

The large majority of LLS have well-bounded, unsaturated detections of
all transitions, or else a detection of \civ ~and \siiv ~without any
detection of \cii ~or \siii, placing them firmly in the top right of
the ionization plane. The ULAS J1342 absorber is qualitatively
distinct from low redshift LLS in its ion ratios.

A portion of the \citet{penprase} metal-poor DLAs could occupy the
same ionization space as the $z=6.84$ absorber, if the true value of
$N_{\rm CIV}$ for the Penprase systems is well below their measured
upper limits (for non-detections) and the true value of $N_{\rm CII}$
is well above their lower limits (from saturation).  This is
consistent with the interpretation advanced in \citet{cooper2019},
where low-ionization absorbers at $z>6$ are early analogs of DLAs,
except that all are younger and more metal-poor.  The distinct space
occupied by the high redshift system in this diagram is a consequence
of slightly more sensitive upper limits on $N_{\rm CIV}$ and $N_{\rm
  SiIV}$, but significantly higher estimates of the column density for
both low ionization species.

\section{Analysis of the Observed Neutral Medium}

At low to intermediate redshifts, DLAs are often interpreted as an
absorption-line manifestation of the multi-phase interstellar medium
described by \citet{mckeeostriker1977} and further refined by
\citet{wolfire1995}.  At $z=6.84$ it is unlikely that interstellar
matter would be gathered into a large, dynamically cold disk. However
the concept of multiple phases in pressure equilibrium and governed by
heating and cooling processes is instructive.

In this paradigm, absorption line measurements can trace four
different states of interstellar matter \citep{2011piim.book.....D}.
First, a {\bf Cold Neutral Medium} (CNM) with $T\sim 100$K exists in
filamentary structures with galactic volume filling factor of a few percent,
but a substantial fraction of the baryonic mass.  Second, a {\bf Warm
  Neutral Medium} (WNM) with $T\sim 7000$ K and lower density occupies
$\sim 20\%$ of the volume, often at the boundary between the
widespread ISM and smaller CNM regions or molecular clouds.  These two
phases are rich in neutral atomic gas and produce the damped \hi
~profiles used to classify DLAs. Third, a {\bf Warm Ionized Medium}
(WIM) with $T\sim 10-15,000$ K---similar to the temperatures inferred
from our Voigt profile fits---and low neutral fraction fills much of
the ISM and is seen in diffuse H$\alpha$, rotation measure of radio
sources, and \civ ~absorption.  These phases are all in pressure
equilibrium with a fourth {\bf Coronal} hot phase, made up of tenuous
($n\sim 10^{-3}$ cm$^{-3}$) ionized gas with $T-10^5-10^7$ K.  The hot
corona fills 30-70\% of the Milky Way's ISM, extends into the wider
circumgalactic medium, and is seen chiefly in \ovi, X-ray line
absorption, or optically thin \hi ~absorption against background
continuum sources.

In nearly all DLAs at $z<6$, one sees evidence for all of these phases
\citep{2003ApJS..147..227P}, as a single sightline can penetrate
regions of CNM, WNM, WIM, and coronal absorption in a single galaxy.
As Figure \ref{fig:ion_ratios} hints, essentially every $z<6$ DLA
with neutral (e.g. \mgi) or singly ionized (e.g. \cii, \siii) metal
lines from the CNM or WNM also exhibits ions from the WIM (\civ,
\siiv) or even \ovi ~\citep{fox2007} that could come from the hot
phase.  This follows naturally from the fact that absorption
probability depends on covering factors or geometric cross-section
\citep{2016ApJ...830...87S}.  The ionized ISM and corona have higher
filling factor than the CNM and WNM, and extend into the CGM. This
explains why highly ionized species are always found in systems
selected by DLA absorption at lower redshift, while the converse is
not true.

In many absorbers at $z>6$ \citep{cooper2019}, and most strikingly in
the system studied here at $z=6.84$, randomly-intercepted metal
absorbers appear to trace some combination of the WNM and CNM, but the
WIM and hot coronal phase are either not present or they can no longer
be traced by rest-UV metal absorption lines.

We speculate three possible explanations for the WIM's disappearance:
\begin{enumerate}
  \item{The WIM does not exist, or has not been heated into
    equilibrium with the CNM and WNM.  This interpretation seems
    unlikely because our detection of \ciistar ~implies some amount of
    local heating in the neutral medium.  If this heating is driven by
    star formation and supernovae, the same processes would also
    naturally yield a hot phase.  In the CGM, much of the heating is
    driven by ubiquitous gravitational processes.}
  \item{The WIM phase exists but has not been enriched with heavy
    elements.  This could affect the cooling properties of the ISM at
    early times.}
  \item{Any photoionization of tenuous gas in the CGM is driven by a
    background radiation field with low ionization parameter and/or
    soft spectrum.}
\end{enumerate}

We evaluate these interpretations by developing a two-phase model of
the neutral medium consistent with all absorption columns plus the
observed heating implied by detection of \ciistar.  We then examine
the expected properties of an unseen interstellar or circumgalactic
hot phase for comparison with observational constraints.

\subsection{\cii$^*$ and Local Sources of Heating and Cooling}

Detections of \ciistar ~require that some fraction of bound electrons
in C$^+$ have been excited from the $^2P_{1/2}$ to the $^2P_{3/2}$
state, at which point they radiate fine structure [\cii] photons at
$\lambda=158\mu$m.  This highly efficient line dominates overall
cooling of the ISM in the Milky Way and local galaxies, and is routinely
observed at high redshift using ALMA.

The emission rate of [\cii] 158$\mu$m radiation (per H atom) can be
measured for C$^+$ atoms whose $^2P_{3/2}$ abundance is measured via
\ciistar ~absorption \citep{pottasch1979}:
\begin{eqnarray}
  l_c & = & h\nu_{ul}A_{ul} N_{\rm CII*}/ N_{\rm HI} \\ & = &
  10^{-26.6} \left({{N_{\rm HI}}\over{10^{20.6}{~\rm
        cm}^{-2}}}\right)^{-1} {~\rm erg}~ {\rm s}^{-1}~ {\rm H}^{-1}.
\end{eqnarray}
In Equation (2) we have substituted our measured value for $N_{\rm C
  II*}$ from Table 1, and scaled to \nhi$=10^{20.6}$.
$A_{ul}=2.36\times10^{-6}$ s$^{-1}$ is the coefficient for spontaneous
decay through [\cii] photon emission, and $h\nu=1.26\times10^{-14}$
erg is the energy of that photon.

\citet{wolfe2003} developed a framework for estimating the star
formation surface density in DLAs where the \ciistar ~abundance can
be measured.  The premise behind this method is that the $^2P_{3/2}$
state must be populated by a combination of (a) radiative excitation
from CMB photons and optical pumping, and (b) local sources of
heating, most of which scale with the local star formation rate (SFR,
hereafter $\dot{\psi}$, measured as a surface density in units of $M_\odot$
yr$^{-1}$ kpc$^{-2}$).

For a fixed SFR, the balance of local heating and cooling results in
an emergent two-phase medium \citep{wolfire1995}, as is seen in the
Milky Way's ISM.  The phase diagram $P(n)$ reveals a small range of
$n$ where two dynamically stable solutions (i.e. $dP/dn>0$) can
co-exist at different density but the same pressure.  These are
conventionally associated with the WNM and CNM.  Gas in these phases
is in pressure, but not thermal equilibrium.  As the SFR increases and
heating is enhanced, the characteristic $n, T$ and $P$ of these stable
solutions for the CNM and WNM increase accordingly.

The same heating/cooling balance calculations used to estimate
$(n,T,P)$ for a given $\dot{\psi}$ also naturally yield a curve of
[CII] emissivity versus density $l_c(n)$, since the 158 $\mu$m line
must be accounted in any budget of ISM cooling.  The emissivity varies
with both metallicity and dust depletion, both of which are observable
in DLAs.

This leads to the following iterative procedure to find the SFR
surface density $\dot{\psi}$ in a DLA where \ciistar ~is detected:
\begin{enumerate}
  \item{Calculate observational bounds on $l_c$ using measurements of
    $N_{\rm CII*}$, an assumption for \nhi, and Equation (1).}
  \item{Specify an initial guess for $\dot{\psi}$, and generate a
    phase curve $P(n)$ for that balance of heating and cooling
    (described below).}
  \item{From the phase curve, evaluate the characteristic stable
    densities of the two phase medium $n_{\rm WNM}$ and $n_{\rm CNM}$, for
    that choice of $\dot{\psi}$.}
  \item{Using cooling estimates from the same calculation and
    additional estimates of photo-excitation of the $^2P_{3/2}$ state,
    extract a second model curve of [CII] emissivity versus density
    $l_c(n)$.}
  \item{Evaluate the model emissivity curve ($l_c(n)$) at $n=n_{\rm WNM}$
    and $n=n_{\rm CNM}$ found in Step 3.  These are the two
    possible/stable values that $l_c$ could take for this choice of
    $\dot{\psi}$.}
  \item{Compare the model $l_c(n_{\rm CNM})$ and $l_c(n_{\rm WNM})$ to
    the observed value from Step (1).  If the model underpredicts
    (exceeds) the data, increase (decrease) the model's $\dot{\psi}$
    and repeat from (2) until convergence.  In general, the WNM will
    match the data for higher values of $\dot{\psi}$ and lower $n$
    than the CNM, because [\cii] has a lower fractional contribution
    to the cooling budget in the WNM.}
\end{enumerate}

Using this method, \citet{wolfe2003} calculated star formation surface
densities for $\sim 30$ DLAs, with average values of $\dot{\psi}\sim
10^{-2.2}$ $M_\odot$ yr$^{-1}$ kpc$^{-2}$ if the absorption is from a
CNM and $\dot{\psi}\sim 10^{-1.3}$ $M_\odot$ yr$^{-1}$ kpc$^{-2}$ for
a WNM.  For comparison, the Milky Way has $\dot{\psi}\sim 10^{-2.4}$
$M_\odot$ yr$^{-1}$ kpc$^{-2}$ \citep{1998ARA&A..36..189K} and Lyman
Break Galaxies have $>1$ $M_\odot$ yr$^{-1}$ kpc$^{-2}$ \citep{2001ApJ...554..981P}.

\begin{figure*}
  \epsscale{1.0}
  \hspace*{-0.5cm}\plotone{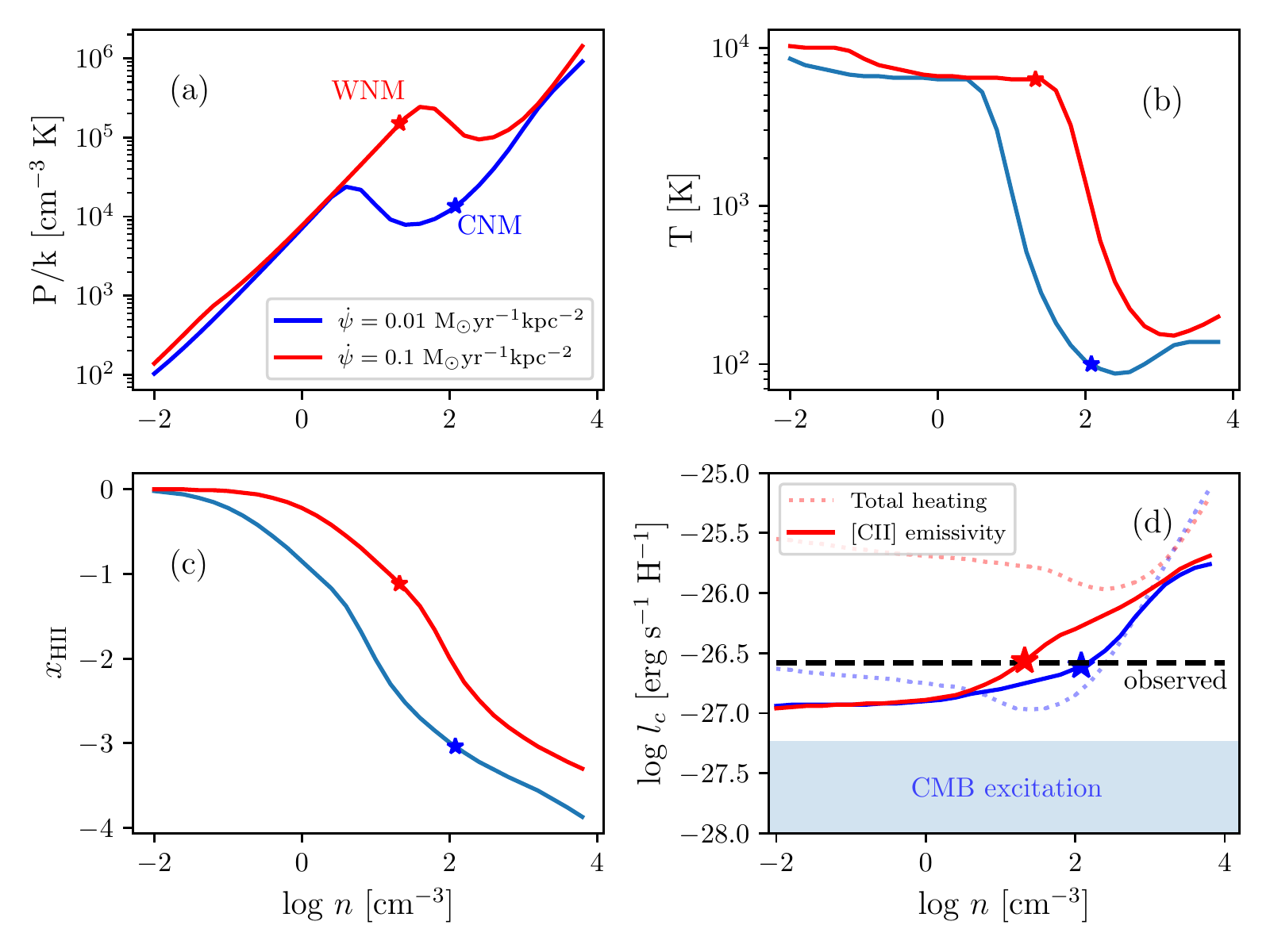}
  \caption{Two-phase model for the neutral ISM at $z=6.84$, assuming
    the cosmic ray ionization rates from \citet{wolfe2003}.  (a) Phase
    curve of pressure vs. density, for star formation densities as
    indicated in the legend.  For each model, at the inflection pair
    in the $P/k$ curve, gas at two different densities can exist at the
    same pressure, leading to a two-phase medium provided $dP/dn>0$
    for stability. For the red curve, a warm neutral medium (left
    branch) matches the derived [\cii] emissivity, whereas a cold
    medium (right branch) matches for the blue curve.  Panel (b) shows
    the associated temperature profiles, with starred points
    indicating the solution matching this absorber.  Panel (c)
    indicates the (logarithmic) ionized fraction and Panel (d) the [\cii]
    emissivity.  Solutions occur at the density $n$ where these curves
    both intersect the observational constraint (indicated with a
    black dashed line) and yield a stable two-phase medium for that
    SFR.  Note that the two starred solutions are {\em not} in
    pressure equilibrium with each other.  Rather, the blue star
    represents a unique model where a CNM can produce the observed
    amount of \ciistar, and is in pressure equilibrium with a warm
    phase not seen in absorption.  The red star represents a model
    where a WNM can produce the observed \ciistar absorption, and is
    in equilibrium with a CNM that is not intersected by the
    sightline.}
  \label{fig:phase_diagrams}
\end{figure*}

\subsection{{\tt cloudy} Implementation of Two-Phase ISM Model}

We have built a revised version of Wolfe's two-phase model to examine
heating and cooling in the neutral medium of our $z=6.84$ absorber.
Using {\tt cloudy}, we simulated a plane-parallel slab of \nhi$=10^{20.6}$
cm$^{-2}$ illuminated externally by a metagalactic background
radiation field.  The external field's normalization and spectral
shape are specified as in \citet{cafg_uvbg} for the absorber redshift.
We also added a blackbody spectrum of CMB radiation for $z=6.84$, as
the CMB becomes a non-negligible source of photo-excitation of the
\ciistar ~fine structure line at low density.

Although we included an external radiation environment, heating of the
neutral medium is in fact dominated by local processes, with the
following four main contributors, all of which scale with the SFR.

\subsubsection{Photoelectric Heating of Dust Grains}
Far-UV radiation (6-13.6 eV) from massive stars can eject
photoelectrons from dust grains, heating the ambient medium via
Coulomb interactions with surrounding electrons. This heating rate
depends on both the local abundance of dust grains, and the strength
of the FUV radiation field which is presumed to be $\propto
\dot{\psi}$.

{\tt cloudy} implements dust in its solution via the {\tt grains}
command, which inserts a mixture of graphite and silicate grains with
size distribution appropriate to the Milky Way's ISM.  However our
absorber's dust abundance should be much lower than that of the Galaxy
on account of its lower heavy element abundances.  We assumed that the
grain abundance scales linearly with gas phase metallicity, and then
applied a modest correction for depletion.

For the model specification of \nhi$=10^{20.6}$, a gas phase metal
abundance of [Si/H]$=-2.2$ yields the observed $N_{\rm Si II}$, and we
scale dust grains downward by the same factor.  An additional downward
correction of $-0.43$ dex is made to the grain abundance to account
for possible depletion as traced by the [Fe/Si]$=-0.2$ ratio, following
Equation 7 of \citet{wolfe2003}.

We deviated slightly from Wolfe's method of modeling the interstellar
radiation field.  Their analysis scales dust heating with $G_0$, the
mean local intensity between 6-13.6 eV \citep{habing1968}, correcting
for different (and unknown) DLA geometries.  Here we use {\tt
  cloudy}'s {\tt Table ISM} command to insert a spectral model of the
ISM radiation field from \citet{black1987}.  We renormalize the
local radiation field's intensity according to the selected star
formation rate, $(\dot{\psi}/10^{-2.4} ~{\rm M}_\odot {\rm
  yr}^{-1}{\rm kpc}^{-2}$), where the normalization factor is again
matched to the star formation surface density of the Milky Way.

\subsubsection{Cosmic Ray Heating}

We introduce cosmic ray heating using the {\tt cosmic rays background}
command in {\tt cloudy}.  Estimates of the Milky Way's primary cosmic
ray ionization rate are evolving and have been updated since
publication of \citet{wolfe2003}. We therefore perform two
calculations of the SFR. The first is for consistency with
\cite{wolfe2003}, and assumes a primary cosmic ray ionization rate of
$1.8\times 10^{-17}$ s$^{-1}$, as in \citet{wolfire1995}.  The second
is the default now assumed for {\tt cloudy}, with a primary CR ionization
rate of $2\times 10^{-16}$ s$^{-1}$
\citep{indriolo2007,indriolo2012}---over $10\times$ higher.

As with the interstellar FUV radiation field, we scale the cosmic ray
ionization rate in units of star formation rate relative to the Milky
Way
 since supernovae are thought to be a significant source
of galactic primary cosmic rays.

Because the \citet{indriolo2007} ionization rates are higher than
those of \citep{wolfire1995}, they generate more heating per unit of
SFR. Consequently these models require lower SFR to achieve levels of
\ciistar ~detected in high-redshift DLAs. Cosmic ray heating is a
non-negligible---and often the dominant---fraction of the total
heating budget in all of the equilibrium solutions found below.

\subsubsection{X-Rays, $H_2$, and Photoionization}

Heating from X-rays and photoionization of \ci ~is treated
self-consistently by {\tt cloudy}, given the input radiation spectrum from
local sources and the extragalactic background.

Heating from collisional de-excitation of vibrational energy from
molecular H$_2$ is also included self-consistently by {\tt cloudy}.  Our
simulations show that this term becomes important at high densities,
above $\gtrsim 1000$ cm$^{-3}$.

\subsubsection{Grid Parameters}

We executed a grid of models with the above heating inputs at fixed
points in density and temperature, ranging from $-2 < n < +4$
cm$^{-3}$ in steps of 0.2 dex, and $1<\log T<5$ K in steps of 0.2 dex.
For each choice of $(n,T)$, we had {\tt cloudy} output the total heating and
cooling rates, the pressure, the \hi ~ionization fraction, and the
total [\cii] 158 $\mu$m emissivity (from both cooling and
photo-excitation). To fix the grid temperature in each run by fiat, we
turned off {\tt cloudy}'s temperature stopping criterion, and instead set
the stopping condition to match $N_{\rm HI}$.

For each value of $n$, there is a single value of $T$ for which the
heating and cooling rates balance.  For each grid step in $n$ we
interpolated the output heating and cooling curves to determine this
equilibrium $T$, reducing these state variables to a one-dimensional
locus.

\begin{deluxetable*}{llcc|cc|l}
  \centerwidetable
  \tablewidth{0pt}
\tablecaption{Two-Phase ISM Model Parameters}
\label{tab:cloudy_results}
\tablecolumns{5}
\tablehead{\colhead{} & \colhead{} & \multicolumn2c{{\tt CLOUDY} cosmic rays} & \multicolumn2c{\citet{wolfe2003} cosmic rays} & \colhead{}\\
\colhead{} & \colhead{} & \colhead{WNM} & \colhead{CNM} & \colhead{WNM} & \colhead{CNM} & \colhead{Observed}}
\startdata
$\log\dot{\psi}$ & (M$_\odot$ yr$^{-1}$ kpc$^{-2}$) & -1.97 & -2.55 & -1.03 & -1.95 \\
$\log n$ & (cm$^{-3}$)             & 1.26 & 2.82 & 1.32 & 2.08 \\
$\log T$ & (K)                    & 3.8 & 1.66  & 3.8 & 2.0 \\
$P/k$ & (cm$^{-3}$ K)              & 5.18 & 4.53 & 5.18 & 4.13 \\
\hline
$N_{\rm CI}$ & (cm$^{-2}$)          & 11.68 & \textcolor{red}{13.34} & 11.12 & 12.44 & $<$12.85\\
$N_{\rm CII}$ & (cm$^{-2}$)          & 14.99 & 14.98 & 15.12 & 14.98 & 15.09 [14.86,15.54]\\
$N_{\rm CIV}$ & (cm$^{-2}$)          & 3.44 & \nodata & 9.37 & \nodata & $<$12.50\\
$N_{\rm CII*}$ & (cm$^{-2}$)         & 13.81 & 13.73 & 13.76 & 13.96 & 13.55 [13.47,13.62] \\
$N_{\rm SiII}$ & (cm$^{-2}$)         & 14.15 & 14.14 & 14.17 & 14.14 & 14.09 [14.00,14.20]\\
$N_{\rm SiIV}$ & (cm$^{-2}$)         & 7.78 & \nodata & 9.60 & \nodata & $<$11.85 \\
$N_{\rm MgI}$ & (cm$^{-2}$)          & 11.52 & \textcolor{red}{12.69} & 11.12 & \textcolor{blue}{11.88} & $<$11.72\\
$N_{\rm MgII}$ & (cm$^{-2}$)         & 14.14 & 14.12 & 14.15 & 14.14 & 13.90 [13.73,14.20]\\
$N_{\rm FeII}$ & (cm$^{-2}$)         & 13.85 & 13.85 & 13.87 & 13.85 & 13.86 [13.77,14.01]\\
$N_{\rm AlII}$ & (cm$^{-2}$)         & 12.47 & 12.47 & 12.48 & 12.46 & 12.44 [12.38,12.50]\\
$N_{\rm AlIII}$ & (cm$^{-2}$)         & 10.11 & 8.80 & 10.86 & 9.73 & $<$11.82\\
\enddata
\end{deluxetable*}

\subsection{Model Results}

Figure \ref{fig:phase_diagrams}(a) shows the resulting phase curves of
$P(n)/k$, as well as (b) $T(n)$ and (c) the ionization fraction
$x(n)$, for the cosmic ray normalization used by \citet{wolfe2003}.  A
stable two-phase medium exists at the inflection of each $P(n)$ curve,
between the local maximum $P_{max}$ and minimum $P_{min}$.  In this
region the density may take on any value, provided $dP/dn>0$ for
stability.  By convention \citep{wolfire1995} we calculate the WNM and
CNM densities at the geometric mean between the maximum and minimum
pressure $P_{eq}=\sqrt{P_{min}P_{max}}$, for each model (i.e. red or
blue curves)

In Panel (d), we show the total heating rate (dashed line), and the
[\cii] 158 $\mu$m emission rate (i.e. the model $l_c$) per H atom
(solid line).  The observed $l_c$ is shown with a horizontal line,
because the density $n$ is not specified by observations and must be
inferred by model comparison.  The intersection of the dashed black
data line and the solid red/blue model lines in panel (d) correspond
to $(n,P,T)$ triplets for a given star formation rate, where the
neutral ISM is in two phase pressure equilibrium {\em and} the [\cii]
emission matches observations.  The model parameters corresponding to
these solutions are shown in Table 3.

On each curve, we have highlighted with a starred point the CNM or WNM
solution that matches the inferred \cii ~cooling rate, $l_c$.  Because
this match occurs at a different value of $\dot{\psi}$ for the CNM and
WNM, the two solutions are shown with blue and red curves,
respectively.  The CNM solution converges on $n=120$
cm$^{-3}$ ($\log n = 2.08$), and $T=100$ K ($\log T = 2.0$).

The star formation surface density that leads to a matching CNM
in this model is $\dot{\psi}=10^{-1.95} ~{\rm M}_\odot ~{\rm yr}^{-1}~{\rm kpc}^{-2}$,
about $3\times$ higher than the Milky Way but lower than
starburst galaxies.

The corresponding WNM solution has $n=21$ cm$^{-3}$ ($\log n = 1.32$),
$T=6300$ K ($\log T = 3.8$), and $\dot{\psi}=10^{-1.03} ~{\rm M}_\odot
~{\rm yr}^{-1}~{\rm kpc}^{-2})$, roughly $10\times$ higher than the CNM
solution and $30\times$ higher than the Milky Way.

For comparison we also show in Table 2 the solution space for the
alternate cosmic ray heating model of \citet{indriolo2007} as
implemented in {\tt cloudy}. Recall that this model has $10\times$
higher CR ionization than Wolfe's for a given SFR.  Stable solutions
exist at lower SFR, since these solutions exist where heating and
cooling balance and the same CR heating is achieved at lower SFR.

For the WNM, the scaling is simple: at $10\times$ lower SFR, the model
converges on the same $(n,T,P)$ triplet as the WNM for the
\citet{wolfe2003} model.  In contrast, the Indriolo CNM solution
occurs at only $3\times$ lower SFR, but much higher density, and lower
$T$. 

\subsection{Comparison with Absorption Columns}

The {\tt cloudy} models developed above produce column density
predictions for all species through the slab, which may be compared
against the observed values.  These are shown at the bottom of Table
2, with the Voigt profile model parameters of Table 1 reproduced in
the rightmost column for ease of comparison.  Recall that these
solutions assume [Fe/H]=-2.2 for the gas phase, \nhi=$10^{20.6}$, and
modest dust depletion as in Equation 7 of \citet{wolfe2003}.  Also, we
have applied slight relative abundance corrections to map the overall
[Fe/H] into each individual element, using the values from Table
\ref{tab:relabund}.

For both models of cosmic ray heating, the WNM solution predicts
column densities consistent with the quoted observational confidence
limits, in every ion. The inferred temperature of $T=6300$ K is also
broadly consistent with the $b$ parameters of our best-fit Voigt
profiles, although the line widths are smaller than our spectral
resolution and hence have large uncertainty.

The CNM absorption columns largely agree for singly ionized species
and above, but show tension between the models and observations for
the neutral species \mgi ~and \ci.  The disagreement is minor and shown
with blue text in Table 2 for the \citet{wolfe2003} model of cosmic
ray heating, but the disagreement is large for the
\citet{indriolo2007} cosmic ray model, and is shown with red text.

This tension arises because the cosmic ray heating becomes so
efficient that very little star formation is needed to maintain
thermal balance at high gas densities.  With few massive stars, the
FUV radiation field at 0.5-0.8 Rydbergs (the ionization energies of
\ci ~and \mgi) is correspondingly weak, resulting in a higher neutral
fraction and correspondingly larger column densities that should be
detected for these species.  Our non-detections of \mgi ~and \ci
~disfavor solutions where the absorption arises from a CNM with
efficient CR heating.

The observations are fully consistent with models of a two-phase
medium where the absorption arises in a WNM of abundance [Fe/H]=-2.2,
SFR surface density modestly higher than the Milky Way, $T\sim 6000$
K, and only minor levels of dust depletion.  CNM solutions with higher
density and lower temperature are marginally consistent with the data,
but only if the coupling between star formation and cosmic ray heating
is less efficient than indicated by the most recent estimates for the
Milky Way.

\subsection{Effect of Varying \nhi}

The {\tt cloudy} models developed above all assume \nhi$=10^{20.6}$
cm$^{-2}$, but this value cannot be constrained by the spectral data.
Here we examine whether the conclusions reached above are robust
if the value of \nhi ~changes.

Suppose that \nhi ~increases by one dex. By Equation 1 the [\cii]
emission rate per H atom will also decrease by 1 dex, since there are
10$\times$ more H atoms per C atom.  In Figure
\ref{fig:phase_diagrams}(d), this corresponds to a reduction of
horizontal black dashed line, toward the region where CMB excitation
becomes a factor in generating [\cii] emissivity.

However the CMB excitation rate per H atom is {\em also} directly
proportional to the carbon abundance, and therefore inversely
proportional to \nhi.  So, even though $l_c$ decreases as \nhi ~grows,
the CMB excitation rate decreases by the same amount.  Put another
way, the observed \ciistar ~column requires more excitation of the
$^2P_{3/2}$ state than the CMB alone can provide even at $z=6.84$, no
matter what value is chosen for \nhi.

The total heating rate does not change as \nhi ~is varied, because
heating is dominated by cosmic rays, which scale with SFR but not
\nhi.  However as \nhi ~increases the heating rate per H atom will
decrease in direct proportion, causing the dotted red/blue lines to
move downward in Figure \ref{fig:phase_diagrams}(d) by the same amount
as the CMB excitation.

The total cooling rate is dominated by [\cii] fine structure emission
and is therefore constrained by the directly observed \ciistar
~column, rather than \nhi.  Again, because the emissivity (solid)
curves in Figure \ref{fig:phase_diagrams}(d) are [\cii] emissivity per
H atom, an increase in \nhi ~lowers these curves by an identical
amount.

Synthesizing all of these factors: varying the \hi ~column density
shifts all of the curves in \ref{fig:phase_diagrams}(d) up or down by
the same amount, so the intersection of the model and observed [\cii]
emissivity (starred points) falls at the same value of $\log n$, and
therefore corresponds to the same value for $T,P$ and the WNM and CNM
solutions.  There may be subtle effects if dust depletion patterns
change with metallicity, or if collisional de-excitation of
ro-vibrational transitions of H$_2$ begin to manifest at high $n$ in
the CNM.  However the basic features of the model, including enhanced
excitation above the CMB baseline, and a stable two-phase solution at
similar points of $(n,T,P)$, do not change with \nhi ~so long as it
remains in the damped Lya or neutral state.

\subsection{Order-of-Magnitude Estimates of [\cii] Luminosity}

The \ciistar ~column density is a direct measure of the surface
density of atoms in the excited fine structure state.  Together with
the spontaneous decay rate, this can be used to project the [\cii] 158
$\mu$m luminosity, given strong assumptions about geometry:

\begin{equation}
  L_{[\rm CII]} = (h\nu_{ul}A_{ul}N_{\rm CII*}) \times 4\pi S_{\rm [CII]}.
\end{equation}

Here, $S_{\rm [CII]}$ is the integrated surface area of the [\cii]
emitting regions, which is model dependent.  If we assume a [\cii]
half-light radius $r_e\sim 3$ kpc measured for $z\sim 6.5$ galaxies
\citep{smit2018}, and set $S_{\rm [CII]}=\pi r_e^2$, this yields a
total estimated luminosity of
\begin{equation}
  L_{\rm [CII]}\sim10^{6.3} ~~ L_\odot
\end{equation}
This estimate is highly speculative, but only partly because of the
strong assumption on the half-light radius\footnote{In the modest
  sample of galaxies for which [\cii] and UV continuum sizes have both
  been measured at $z>5$, there is evidence that $r_e$ is $2-3\times$
  larger in [\cii] emission than in the stellar continuum
  \citep{2020ApJ...900....1F,2019ApJ...881..124M}}. Equally important,
the absorption properties are slightly more consistent with a WNM
(which has a larger volume and covering factor), yet for many galaxies
the total [\cii] emission is dominated by CNM gas in pressure
equilibrium that is simply missed by the quasar absorption path.  If
the emitting region covers only a small fraction of the half light
radius, then Equation 4 could overestimate the [\cii]
luminosity. However if the absorbing gas traces a WNM that is
accompanied by stronger emission from the CNM, then Equation 4 would
underestimate $L_{\rm [CII]}$.

As a separate consistency check, one can start instead with our
estimates of the SFR surface density $\dot{\psi}$, calculate a total
SFR integrated over the galaxy, and then use empirical or model
correlations between total SFR and $L_{\rm [CII]}$ to estimate the 158
$\mu$m luminosity.

If we exclude the CNM model with {\tt cloudy}'s default cosmic ray
ionization (because of tension with \ci ~and \mgi), the other models
all have $\dot{\psi}\sim 0.01-0.1$ M$_\odot$ yr$^{-1}$ kpc$^{-2}$.
Again combining this with the [\cii] half-light radius $r\sim 3$ kpc
\citep{smit2018}, one derives a total star formation rate of order
$0.3-2.0 M_\odot$ yr$^{-1}$. Comparing to the [\cii]-SFR correlations
modeled by \citet[][Equation 10]{lagache2018}, we estimate:
\begin{equation}
  L_{\rm [CII]}=10^{6.13}-10^{6.89} L_\odot
\end{equation}
For EoR galaxies with $\log({\rm SFR})\sim 0$, \citet{finlator2018}
estimate stellar mass of $\log M_*\sim 8.5$ M$_\odot$.

These calculations, while crudely consistent, are meant to be
heuristic guides rather than precise measurements.  They demonstrate
that absorption selection does uncover systems that are locally
heated, yet produce quotidian [\cii] luminosities and stellar
population properties, much less extreme than the galaxies of high SFR
or metallicity observed at high redshift
\citep{2019ApJ...870L..19N,2020Natur.581..269N}.  Such systems are
predicted by numerical simulations but largely missed in all but the
deepest flux-limited blind galaxy surveys.  

\section{Non-Detection of Warm Ionized Gas}

\subsection{No detection of metals in an ionized ISM}

Multiphase models of the Milky Way's ISM include a WIM in pressure
equilibrium with the WNM; this WIM is traced in low surface-brightness
H$\alpha$ emission \citep{WHAM}, rotation measures of radio sources
\citep{dispersion}, and in \civ ~absorption against the spectra of hot
stars \citep{savage1997, sembach2000}.  It is diffused through a large
fraction of the Milky Way's ISM volume with scale height of 1-3 kpc
above the disk.

Warm ionized (\civ) and hot (\ovi) gas have also both been studied in
low-redshift DLAs; they are not precisely coincident in velocity
space with the neutral medium that produces low-ionization lines
\citep{fox2007}.  The hot \ovi ~shows broad absorption widths and is
consistent with a collisionally ionized gas at $T\approx 300,000$ K.
The \civ ~absorption is slightly narrower and could be associated with
either collisionally ionized gas at $T=100,000$ K where the \civ
~collisional fraction peaks, or at slightly lower $T$ with enhancement
from photoionization.  However standard thermal equilibrium
assumptions are not always justified for the warm \civ ~gas.  It can
become unstable to cooling, yet during this cooling it can experience
large departures from ionization equilibrium, such that the \civ
~fraction remains high as the gas cools to stability at $T=10,000$ K
\citep{oppenheimer2013}.

In our multiphase model of the $z=6.84$ absorber, gas at the peak \civ
~ionization temperature of $T=10^5$ K would be in pressure equilibrium
with the WNM ($P/k\sim10^{5.2}$ cm$^{-3}$ K) if it has density
$n=10^{-0.2}=0.6$ cm$^{-3}$.  We ran simple {\tt cloudy} calculations
with these input $(n,T)$, and the WNM model values in Table 2 for the
SFR $\dot{\psi}$, {\tt cloudy}-default cosmic ray heating, including
local+extragalactic background radiation fields.  Because \nhi ~is not
known and the ionized gas may be optically thin at the Lyman limit, we
set the stopping criterion according to cloud thickness rather than
\nhi, using 1 kpc as a reference value.

With these inputs, if we assume the same value of [Fe/H]$=-2.2$ that
was calculated for the WNM and CNM models, {\tt cloudy} calculates a
WIM ionized fraction of $\log x=-4.8$, and column densities:
\begin{eqnarray}
  N_{\rm HI}&=&10^{16.51} \left({{\Delta L}\over{1 ~{\rm kpc}}}\right) \\
  N_{\rm CIV} &=& 10^{14.57}\left({{\Delta L}\over{1 ~{\rm kpc}}}\right)\\
  N_{\rm SiIV}&=&10^{13.41}\left({{\Delta L}\over{1 ~{\rm kpc}}}\right).
\end{eqnarray}
While this value for \hi ~is plausible, the predicted \civ ~column
exceeds our observational bounds by 2.1 dex, and the \siiv ~limits by
1.6 dex.

To bring this WIM model \civ ~column into alignment with measurements,
either its length scale must be reduced by a factor of $10^{-2.1}$ to
$\sim 8$ pc, or the heavy element abundance must be reduced by 2.1 dex
to [C/H]$<-4.2$ (95\% confidence).  For reference, the cloud depth
required to reach DLA column densities for the WNM {\tt cloudy} model
(i.e. Table 2) is $\sim 20$ pc. If the WIM is enriched to the same
abundance as the WNM, then it is over a $2-3\times$ smaller length
scale and therefore its metals cannot be considered widely mixed into
the ISM.  Alternatively, any heavy elements mixed spatially into in
the WIM on kpc or larger scales must be at $\sim 100\times$ lower
abundance than the neutral medium where stars form.  In some ways this
represents an inversion of the prevailing paradigm of ISM and CGM
absorption, where small neutral condensates are embedded in an ionized
medium occupying a larger fraction of the volume
\citep{2016ApJ...830...87S}.  We are not arguing here that the ionized
phase does not exists, only that it may not yet manifest in metal
absorption because of low enrichment.

\begin{figure*}
  \epsscale{0.75}
  \hspace*{-0.5cm}\plotone{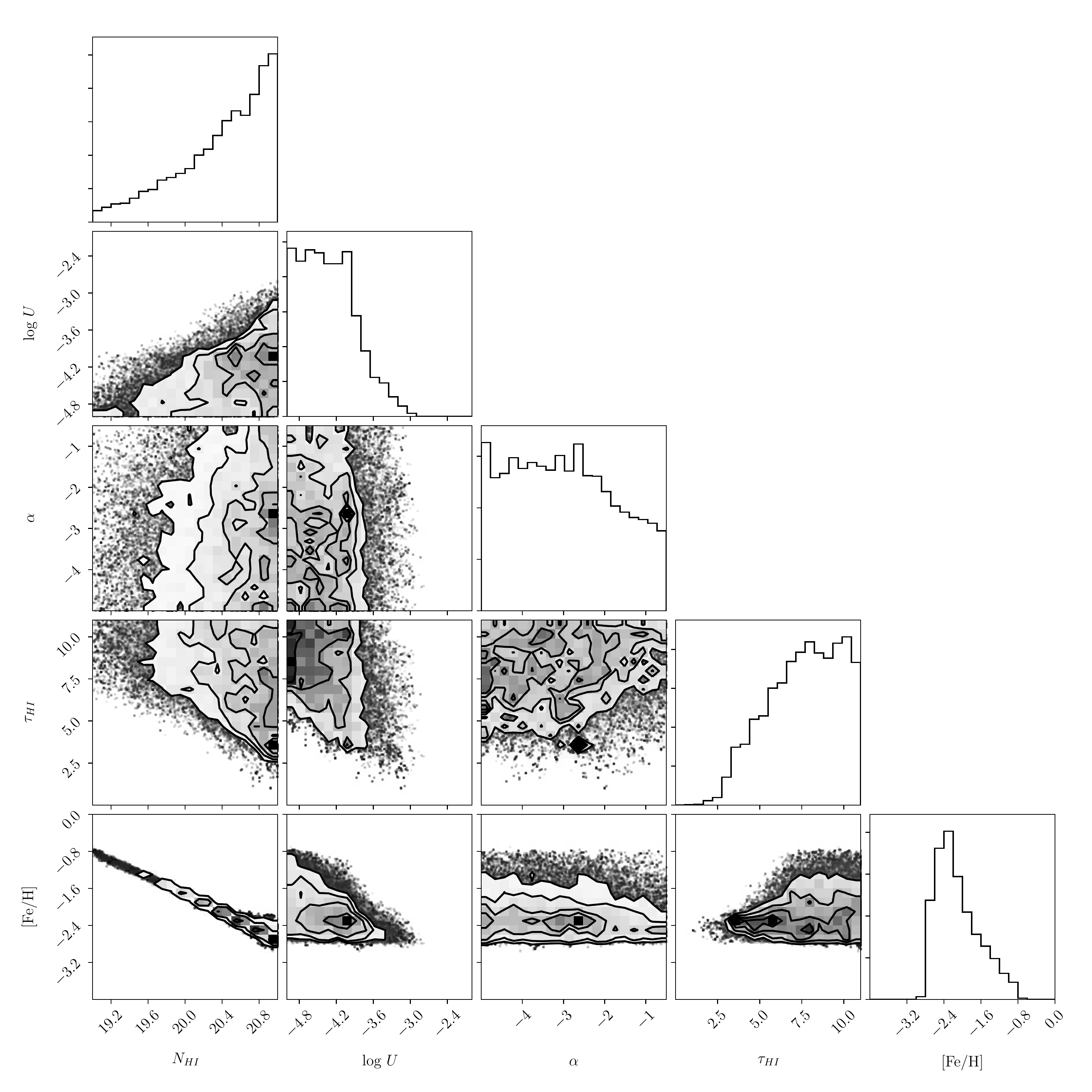}
  \caption{Corner plot of posterior distributions for the 5-parameter
    model discussed in Section \ref{sec:cgm_analysis}.  The narrow
    line in the \nhi ~vs. [Fe/H] plot traces out abundance values for
    a neutral medium, and the marginalized posterior of \nhi
    ~illustrates how DLA-like column densities are favored by the
    measured ratios of metal ions.  The non-detection of \civ
    ~requires a weak ionization parameter $U$.  A large break is
    required at the Lyman limit ($\tau_{\rm HI}$), so that there are
    enough $E<13.6$ eV photons to ionize \mgi, but few photons to
    produce \civ ~and \siiv.  At the large \hi ~columns that are
    favored, the system is self-shielded from hard-UV photons, so an
    unusually soft ionizing spectrum is not necessarily required.}
  \label{fig:uvbg_posterior}
\end{figure*}

A WIM with this $(n,T)$ combination and abundances is not in thermal
equilibrium. The {\tt cloudy}-calculated cooling rate exceeds the
heating rate, meaning that \civ ~in a $T=10^5$ K WIM should cool on
short timescales compared to the Hubble time.  Non-equilibrium effects
may enhance the \civ ~fraction relative to its photoionization
equilibrium value as the cooling gas approaches $T\sim 10,000$ K from
above.

The non-detections of \civ ~and \siiv ~imply that if this galaxy has a
warm ionized component to its ISM as seen in the Milky Way and all
lower redshift galaxies, then this WIM has not yet been enriched by
heavy elements at $z=6.84$.  Or, at least the degree of enrichment is
substantially lower in abundance or occupies a smaller volume than
nearby neutral regions.

\subsection{No Detection of an Ionized Circumgalactic Medium}
\label{sec:cgm_analysis}

The multiphase models developed above are motivated by observations of
interstellar matter in the Milky Way and other local galaxies, within
their central, dynamically cold disk.  Yet at lower redshift where
host galaxies are more easily surveyed, most absorption systems and
even many DLAs are located at large impact parameter ($10-100$ kpc)
from the central galaxy and its associated ISM
\citep{2020arXiv201011958B,2017Sci...355.1285N,2014MNRAS.445..225C,2017ARA&A..55..389T}.
If the $z=6.84$ absorption traces the extended tenuous CGM of a
protogalaxy, a similar set of {\tt cloudy} models can be used to
constrain its ionization conditions.

Suppose we wish to generate a single-phase CGM absorption model
consistent with strong detection of all low-ionization species, but
sensitive non-detection of all conventional highly-ionized lines.
Such a model requires some combination of a low ionization parameter
$\log U$, self-shielding from a large \hi ~column density, or a very
soft spectral shape of the ionizing background radiation field.

To explore the trade space between these options, we ran a
5-dimensional grid of {\tt cloudy} models varying the following
parameters:
\begin{itemize}
\item{\nhi: The neutral \hi ~column is not an observable, and must be
  treated as a model variable.  We set a flat prior distribution in
  the logarithm with $19<\log N_{\rm HI}<21$ cm$^{-2}$.}
\item{Ionization parameter: We varied $\log U$ from $-5$ to $-2$ with
  a flat prior in log space.}
\item{Spectral slope: this parameter $\alpha$ sets the hardness of the
  UV background field $f_\nu\propto \nu^{\alpha}$ at energies above 1
  Rydberg.  We varied it between very hard values of $\alpha=0$ to
  very soft values of $\alpha=-6$, with flat prior.}
\item{Spectral break $\tau_{\rm HI}$ at the Lyman edge: This is a
  feature of the UV background spectrum, specifying the strength of a
  decrement across the 1 Rydberg break.  We sampled uniformly between
  $0<\tau_{\rm HI}<10$.}
\item{Metallicity [Fe/H]: this parameter was varied between
  $-5<$[Fe/H]$<0$.}
\end{itemize}
At each point on the grid, {\tt cloudy} runs a simulation with
stopping condition set on the specified \hi ~column density,
outputting column density predictions for every observed ion.  We read
these into a set of 5-d interpolation tables to calculate rapid column
density predictions for any vector of model input parameters.

We explored the parameter space of these five variables with an {\tt
  emcee} MCMC walker, calculating the likelihood at each step by
comparing the model outputs with the measured column densities or
upper limits for each ion, as described in
\citet{2015MNRAS.446...18C}.  A corner plot of the output is shown in
Figure \ref{fig:uvbg_posterior}.  Because we do not measure \nhi ~and
only have upper limits on \civ, \siiv, and \mgi, the MCMC walker does
not converge on two-sided bounds for most model parameters.  Instead,
it delineates one-sided or diagonal bounds in each plane corresponding
to allowed or disallowed regions of the parameter space.

A line of degeneracy in the [Fe/H] vs. \nhi ~posterior traces out the
solution locus for a neutral medium with the observed metal-line
columns, but the marginalized histogram of \nhi ~shows a clear
preference for high \hi ~column density values in the DLA range ($\log
N_{\rm HI}>20.3$ by convention).  The heavy element abundance is
similar to the values used for the two-phase ISM model above (recall
this model assumed \nhi$=10^{20.6}$).  Put another way, even when the
walker explores a large solution space, the posterior ends up favoring
a solution with \hi ~column density characteristic of a DLA and
abundance slightly below $1/100$ solar, rather than a more tenuous and
ionized CGM.

The ionization parameter posterior favors very small values
$U<10^{-4}$ which flow from the low \civ/\cii ~and \siiv/\siii ~ratios.
Importantly, for solutions with lower \nhi ~and higher [Fe/H] (which
are more likely to require ionization corrections), the smallest
values of $U$ are required to prevent overproduction of \civ.

Likewise, while all solutions require a strong break at 1 Rydberg of
$\tau_{\rm HI}\gtrsim 3$, the lowest values of \nhi ~require the
largest spectral breaks.  For \nhi$<10^{20}$, breaks of $\tau_{\rm
  HI}\gtrsim 6.5$ or more are needed.  This is primarily to reproduce
the large observed \mgii/\mgi ~ratio.  Large breaks at 1 Rydberg
ensure that there are enough soft-UV (0.56 Ryd) photons to ionize \mgi
~to \mgii, but not too many hard-UV photons that would ionize \cii ~to
\ciii ~(1.79 Ryd), or \ciii ~to \civ ~(3.52 Ryd).

This exercise demonstrates that even if we search a wide parameter
space for single-phase solutions corresponding to more tenuous
circumgalactic gas, an MCMC walker will still converge on models
resembling a metal-poor but not chemically pristine DLA.  In fact the
inputs are very similar to the multiphase ISM model described
earlier. Such systems could exist in the CGM of a protogalaxy, but
would need to support some amount of local star formation to reproduce
the observed \ciistar.

\section{Discussion}

The models outlined above shape our physical intuition about the
environment of a star forming region randomly selected by intervening
absorption cross section, when the Hubble time was just 794 Myr.  This
is among the first detailed characterizations of a non-extreme
astrophysical environment in the epoch of reionization.

Strong absorption from singly ionized species such as \cii, \siii,
\alii, and \feii ~requires a medium with a substantial \hi ~neutral
fraction, yet the lack of \mgi ~and \ci ~implies some degree of
ionization from the FUV radiation field of massive stars.

A confident detection of the \ciistar ~fine structure line constrains
heating of the ISM by radiation and cosmic rays associated with those
stellar populations and their associated supernovae. The inferred star
formation surface density depends on whether the absorption traces a
warm or cold neutral medium.  However the derived SFR values are
within a factor of three of what is seen in the Milky Way, and do not
require exotic stellar populations, modifications to the interstellar
radiation field or cosmic ray physics.

These two-phase ISM models---tuned to achieve pressure balance,
heating/cooling equilibrium, and the [\cii] emissivity---also correctly
reproduce the observed column densities of all absorption species
(except for CNM models with efficient cosmic ray heating, which
overproduce \mgi).  According to the {\tt cloudy} outputs, neutral gas
in this phase will never produce observable high-ionization absorption
from \civ ~or \siiv.

The heavy element abundance depends upon one's choice of \nhi ~(which
is unconstrained by observations). For \nhi$=10^{20.6}$ (a modest DLA),
an abundance of [Fe/H]$=-2.2$ yields matching column densities; larger
choices of \nhi ~lead to correspondingly smaller metallicity. This
abundance is slightly lower than the median DLA metallicity at $z<4$,
but not a strong outlier.  Apparently rapid enrichment was already
well underway in the neutral medium of this system, within its first
few hundred Myr.

At lower redshifts, the \cii ~and \siii ~which trace the optically
thick neutral medium are always accompanied by strong \civ.  This
either arises in a warm-ionized phase of the ISM near the collisional
ionization optimum for \civ ~($T=10^5$ K), or else in lower density
circumgalactic gas photoionized by the ambient background field.  This
high-ionization signature is completely absent in the system studied
here, with stronger constraints than heave been measured for other
DLAs at any redshift.

If a warm ionized phase exists in this system at $T\sim 10^5$ K, the
non-detection of \civ ~places a degenerate constraint on its
absorption pathlength and carbon abundance.  For $\Delta L=1$ kpc (the
scale height of the Milky Way's WIM), one obtains [C/H]$<-4.6$ ---
indicating that metals produced in the neutral regions have not yet mixed
throughout the ionized ISM.

The relative abundances of heavy elements do not show any noteworthy
deviations from the Solar pattern, adding to the evidence that
enrichment was driven by conventional Population II stars, rather than
a top-heavy or Population III initial mass function.

It is possible to make order-of-magnitude estimates for the stellar
mass and [\cii] luminosity of the associated stellar population, under
a strong assumption that the [\cii] emissivity per kpc$^{-2}$ may be
integrated over an area similar to the [\cii] half light radius of
other $z\sim 6.5$ objects (which is larger than the UV/optical half
light radius).  Using two different methodologies we estimate [\cii]
luminosity of $10^{6.1}-10^{6.9}L_\odot$.  Integrating the SFR surface density over
the [\cii] half-light radius (assumed as 3 kpc), one obtains a total SFR of $O(1)$ M$_\odot$
yr$^{-1}$.  Simulated EoR galaxies with this SFR have stellar masses
in the vicinity of $M\sim10^{8.5}M_\odot$, similar to the Small
Magellanic Cloud.

\section{Conclusions}

We have analyzed in detail the most distant quasar absorption system
now known, at $z=6.84$ in spectra of the $z=7.54$ quasar ULAS
J1342+0928.  Using data obtained over 33 combined integration hours
with Magellan/FIRE and VLT/XShooter, we fit Voigt profile column
densities to 12 different ionic species.  Besides being the most
distant quasar absorber, this is also the most extreme example of the
low-ionization systems studied by \citet{cooper2019}, where large columns
of \cii, \mgii, and \siii ~are seen without any corresponding
absorption from highly ionized \civ ~ or \siiv.

Our findings may be summarized as follows:
\begin{enumerate}

\item{Strong detections of \mgii, \cii, \siii, \feii, \alii, and
  especially \ciistar ~suggest that this absorber would classify as a
  Damped Lyman Alpha (DLA) system, even though we cannot measure \nhi
  ~because of blending with the \lya forest.  Assuming a fiducial
  DLA column density of \nhi$=10^{20.6}$ cm$^{-3}$, a heavy element
  abundance of [Fe/H]$=-2.2$ reproduces the observed metal line column
  densities, modulo small corrections for relative abundances.  Larger
  assumed values of \nhi ~require proportionally lower metallicities.}
\item{The relative abundances of [C/Fe], [Si/Fe], [Al/Fe] and [Mg/Fe]
  show only modest departures from the Solar value, and are consistent
  with patterns observed in metal-poor DLAs at lower redshift, as well
  as metal-poor stars in the Milky Way's halo.  Just 800 Myr after the
  Big Bang, there is no evidence of enhanced [$\alpha$/Fe] as would be
  expected if yields were dominated by core collapse supernovae, nor
  is there a [C/Fe] enhancement as is seen in extremely metal-poor
  stars.}
\item{This system exhibits the lowest ratios of $N_{\rm CIV}/N_{\rm
    CII}$ and $N_{\rm SiIV}/N_{\rm SiII}$ of any quasar absorption system
  at any redshift.  This results from strong detection of saturated
  singly ionized lines, and sensitive non-detection of the highly
  ionized species.  If there is an ionized phase of the ISM or CGM in
  this environment, it has not yet been enriched detectably with heavy elements.}
\item{A clear detection of the \ciistar ~excited fine structure line
  requires local sources of heating in the neutral medium, as the
  implied excitation rate is $\sim 5\times$ higher than CMB photons
  can provide at this redshift.  Following \citet{wolfe2003} we
  interpret this as evidence for local star formation in proximity
  to the neutral medium.}
\item{We balance heating and cooling functions calculated by {\tt
    cloudy} to develop a two-phase warm/cold model of the neutral
  medium.  These models yield estimates of the SFR that can maintain
  the observed levels of $^2P_{3/2}$ excitation of \cii, and also
  produce estimates of the [\cii] 158 $\mu$m emissivity per square
  kpc.  These simulations rely on evolving local calibrations for the
  cosmic ray primary ionization and heating rates, but yield
  instructive comparisons to the Milky Way.}
\item{The simulations show an emergent two-phase medium if the star
  formation surface density is slightly higher than the Milky Way but
  lower than starburst galaxies.  We compare metal-line column
  densities for both the WNM and CNM solutions to observations,
  finding a slight tension between models and data for the cold
  medium.  This is because the CNM models require very little star
  formation to excite the \ciistar ~fine structure line; the weak UV
  radiation field that results leads to residual neutral \ci ~and \mgi
  ~that are ruled out by non-detection in the data.  We conclude that
  the singly ionized species are most consistent with a warm neutral
  phase of the ISM.}
\item{Under the strong assumption that this [\cii] emissivity can be
  integrated over a 3 kpc [\cii] half light radius---motivated by
  galaxies at similar redshift---we estimate an order-of-magnitude
  total $L_{[\rm CII]}\sim 10^{6.1}-10^{6.9}~L_\odot$.  This level of 
  $L_{[\rm CII]}$ and
  star formation are present in numerically simulated galaxies with
  stellar mass $M_*\sim 10^{8.5}M_\odot$ \citep{finlator2018}, similar
  to or slightly smaller than the Small Magellanic Cloud
  \citep{stanimirovic2004}. }
\item{We ran {\tt cloudy} models of a warm ionized phase in pressure
  equilibrium with the neutral gas, to investigate why we do not
  detect \civ ~and \siiv.  These models yield \civ ~and \siiv ~column
  densities much larger than observed upper limits, if the warm medium
  has a metallicity comparable to the neutral phase, and a pathlength
  of 1 kpc (similar to the WIM scale height in the Milky Way).  This
  discrepancy may be reconciled if the WIM has an abundance of
  $\lesssim\frac{1}{100}$ that of the WNM, or if it has a very small
  absorption path ($\Delta L\lesssim 8$ pc).  If this interpretation
  is correct, then heavy elements present in the WNM have not yet
  mixed into the ionized ISM or CGM as is observed at lower redshifts,
  implying that this system is still at an immature stage of chemical
  enrichment.}
\end{enumerate}

Quasar absorption systems provide our highest-SNR spectral
measurements in the Epoch of Reionization, and correspondingly rich
astrophysical insight about ``normal'' galaxies selected by
straightforward cross section rather than extreme luminosity or star
formation.  Numerical simulations predict that many or most of the
galaxies driving reionization resemble modern-day dwarfs, and will be
challenging to detect even with the James Webb Space Telescope.  Quasar
absorption may be the best way to study such systems in detail, for
the foreseeable future.

Yet analysis of absorption systems is complicated by the usual
challenge of modeling geometry, and at $z>6$ the inability to measure
\nhi ~directly adds another layer of uncertainty.  Nevertheless it is
possible to draw many robust conclusions from metal-line observations
alone, and also to build plausible models which yield more physical
insight so long as their model-dependence is suitably disclosed.

The data alone indicate that within several hundred Myr, this system
has experienced substantial enrichment in gas with a high neutral
fraction, and that the enrichment pattern already is dominated by
conventional stellar populations rather than an exotic, or top-heavy
IMF.  We look forward to future observations of $z>6.5$ absorbers that
are closer to the quasar's emission redshift. In such systems the \oi
~1302\AA ~line can probe both the neutral fraction and the [O/Fe]
ratio, which is enhanced for a Population III-like IMF
\citep{2019MNRAS.487.3363W,2010ApJ...724..341H}.  While \oi
~measurements have been made at $5.8 \lesssim z \lesssim 6.4$ in a
number of proximate DLAs
\citep{2019ApJ...885...59B,2018ApJ...863L..29D} and foreground
absorbers \citep{2011ApJ...735...93B,cooper2019}, it appears that one
needs to search even earlier epochs for clear evidence of anomalous
yields.

A total abundance measurement is impossible without constraints on
\nhi, but for standard DLA column densities the enrichment level is in
the range of $-3<$[Fe/H]$<-2$---i.e. metal-poor but not chemically
pristine or even ultra-metal-poor.  There is strong evidence for local
heating of the neutral ISM.

Ionization models demonstrate that one can reproduce all observables
if the neutral gas resides in the warm component of a two-phase
neutral medium like that of the Milky Way, with star formation surface
density only slightly higher than that of the Galaxy.  Any inferences
about an associated stellar population require aggressive assumptions
about geometry.  However they are at least plausibly consistent with the
interpretation that the absorption is from the ISM or CGM of an
SMC-like galaxy with $M\sim 10^8-10^9 M_\odot$, a star formation rate
of $O(\sim 1) M_\odot$ yr$^{-1}$, and modest [\cii] luminosity.

The disappearance of an enriched, ionized phase of the ISM or CGM
appears to be quite common at $z>6$ and this system as an extreme
example provides and opportunity to investigate the phenomenon in
detail.

\vspace{5mm}
\facilities{Magellan (FIRE) \citep{2013PASP..125..270S}}
\facilities{VLT (XShooter) \citep{xshooter}}


\software{astropy \citep{astropy:2018,astropy:2013}
          Cloudy \citep{2013RMxAA..49..137F}
          emcee \citep{emcee}
          PypeIt \citep{pypeit}
          }



\acknowledgements

The authors gratefully acknowledge contributions from Fabian Walter in
planning and executing the XShooter program, and consulting on
scientific content.

This research made use of Astropy, a community-developed core Python
package for Astronomy \citep{astropy:2013, astropy:2018}. 

ACE acknowledges support by NASA through the NASA Hubble Fellowship
grant $\#$HF2-51434 awarded by the Space Telescope Science Institute,
which is operated by the Association of Universities for Research in
Astronomy, Inc., for NASA, under contract NAS5-26555.

This work was supported by the ERC Advanced Grant 740246 “Cosmic gas.”

Based in part on observations collected at the European Southern
Observatory, Chile, programme IDs 098.B-0537(A) and 0100.A-0898(A).

This paper includes data gathered with the 6.5 meter Magellan
Telescopes located at Las Campanas Observatory, Chile.

\clearpage

\bibliography{hiz}

\end{document}